\documentclass[11pt]{article}
\usepackage[margin=1in]{geometry}
\usepackage{amsmath,amssymb,amsfonts,mathtools}
\usepackage{textcomp}
\usepackage{algorithm}
\usepackage{algpseudocode}

\usepackage{graphicx}
\usepackage{booktabs}
\usepackage{array}
\usepackage{tabularx}
\newcolumntype{L}{>{\raggedright\arraybackslash}X}
\usepackage{tikz}
\usetikzlibrary{arrows.meta,positioning,shapes.geometric,fit}
\usepackage{stfloats}
\usepackage{url}
\usepackage{verbatim}
\usepackage{float}
\usepackage{subfig}
\usepackage{microtype}
\usepackage{xcolor}
\usepackage[numbers,sort&compress]{natbib}
\usepackage{hyperref}
\hypersetup{hidelinks}

\title{Receding Horizon Multi-Agent Deceptive Path Planner}
\author{%
Xubin Fang$^{1}$, Brian M. Sadler$^{2}$, and Rick S. Blum$^{1}$\\[0.6em]
\small
\parbox{0.90\textwidth}{%
\centering
$^{1}$ Department of Electrical and Computer Engineering,\\
Lehigh University, Bethlehem, PA 18015, USA\\[0.4em]
$^{2}$ Oden Institute for Computational Engineering and Sciences,\\
The University of Texas at Austin, Austin, TX 78712, USA\\[0.5em]
xuf220@lehigh.edu;\quad
brian.sadler@ieee.org;\quad
rblum@eecs.lehigh.edu
}%
}

\date{}

\begin{document}
\maketitle

\begin{abstract}
Deceptive path planning enables autonomous agents to obscure their true goals from observers by deviating from an expected optimal path. Prior work largely solves full-horizon, end-to-end optimization for single agents, 
{so planning complexity grows with the path length and size of the environment, and adaptation en route requires a new optimization. 
We propose a unified framework for deceptive path planning onboard the agent, 
computing over short-horizon candidate paths within a receding-horizon loop.
The method processes an agent’s position sensor information and develops a policy for the next trajectory plan. 
By parameterizing a user-defined cost that captures optimal planning and deception 
(and optionally includes constraints, trajectory smoothness, and 
coupling terms between agents), a Boltzmann framework yields stochastic policies that balance the tradeoff between optimal paths and deceptive deviation. Policies are updated locally and do not require learning or training.} The level of deception and adherence to constraints can be dynamically tuned, enabling online adaptation to changes in goals and constraints. This step-by-step tuning opens the door to new forms of dynamic deception. In the multi-agent case, we develop a joint planner that is also tunable and can be applied as desired among all or subsets of agents.
Single- and multi-agent simulation studies demonstrate the flexibility of our approach, maintaining deception while adapting as desired, avoiding the recomputation required by full-horizon methods, and supporting intuitive tuning via a small set of parameters.
\end{abstract}

\vspace{1em}

\section{Introduction}

Path planning is fundamental for multi-agent coordination and control \cite{ref_fregene2005}.
Examples include full coverage path planning \cite{ref_fu2025}, planning on graphs for patrolling, routing, and surveillance \cite{ref_wang2025}, and search \cite{ref_papaioannou2023}.
Time constraints can be incorporated with linear temporal logic \cite{ref_tian2022}, and planned rendezvous can be included for persistent networking and information transfer \cite{ref_yu2020}. 
Local trajectory planning with nearest neighbor communications can provide adaptive collision avoidance~\cite{ref_zhou2019}. 

In this paper, we consider 
deceptive path planning~(DPP),
where agents obscure their true goals from observers. Deceptive paths are characterized by deviation from an optimal (e.g., shortest) path to the intended goal, using exaggeration, ambiguity, or other deceptive measures aligned with false goal states. The adversary is modeled as a passive observer who assumes agents will follow an optimal path. 
DPP is important in adversarial scenarios where an observer may attempt to predict the agent's destination goals.
This applies to a variety of domains, such as resource allocation, human--robot interaction under adversarial observation, and autonomous delivery and logistics in competitive environments~\cite{chen2022,fatemi2024dpp,savas2021}. Much of the existing research on DPP focuses on single-agent cases and generating one deceptive path, assuming a known environment~\cite{chen2022,savas2021,xu2020single,fu2022almost,lv2024optimal,price2023domain,lenhard2025cadetopath,xu2020scalable}. 
The complexity of end-to-end planners grows with the path length and environment size, and requires replanning when the environment, deception type, constraints, or other system parameters change
\cite{fatemi2024dpp,xue2025efficient,chen2024dpp}. 

We present a general receding horizon deceptive path planning framework that can be applied in single- and multi-agent settings and enables dynamic adaptation.  
The method relies on local sensing of position and obstacles and plans the local trajectory or action sequence. 
To model decision-making, we construct a probability mass function (PMF) based on a Boltzmann distribution over candidate trajectories, which is parameterized by a user-specified cost $C(\cdot)$. This PMF defines a stochastic policy from which steps are sampled and sequentially added to the path. A multi-agent formulation is obtained by lifting this distribution to the joint space of agents’ trajectories, and costs can be coupled among groups of agents as desired.
At each iteration, the planner performs a finite-step lookahead. The goals, false goals, constraints, and form of deception can all be updated at each planning iteration. The cost function $C(\cdot)$ and its parameters are updated (if desired), the policy is updated, and a new finite-step lookahead begins. This receding-horizon loop
repeats until the agent reaches its goal.
The approach is simple to implement, allows for easy modification of costs and tuning of the level of deception at each iteration, is robust to local environment dynamics, and generalizes to multiple agents.

{
The main contributions of this paper are: 
\begin{itemize}
    \item  A DPP framework that uses local position information and enables adaptation to a dynamic environment, constraint changes, goal modifications, and moving goals, without end-to-end replanning. The proposed planning framework is simple and could be integrated into standard navigation pipelines.
    \item A flexible Boltzmann-based formulation for local random policy design that enables agents to adapt their deceptive behavior with time-varying intuitive parameters.
    \item A logical extension from single to multi-agent scenarios with coupled costs and/or incentives while allowing individual agent variation in goals and parameter tuning. Agents collaborate by sharing a limited parameter set.
    \item A generalizable approach that can handle various DPP scenarios and extends to other path planning problems by adjusting the cost functions. Cost terms may include incentives that reduce the total cost.
    \item A variety of simulation experiments illustrating the practical utility and adaptability of our framework, enabling new forms of dynamic deception.
\end{itemize}
}

Section~\ref{LR} reviews related work in DPP. In Section~\ref{AL}, we present the  
single-agent case, 
providing the foundation for the multi-agent extension discussed in Section~\ref{MA}. 
Simulation results 
for various single- and multi-agent scenarios
are presented in Section~\ref{ER} to demonstrate the flexibility and effectiveness of our framework. 
Finally, Section~\ref{CL} concludes the paper and outlines directions for future research.

\section{Related Work}
\label{LR}
Deceptive path planning (DPP) has roots in
early studies of deceptive
motion in human--robot interaction (HRI)~\cite{dragan2015deceptive} and in
multi-agent reconnaissance with deception strategies~\cite{root2005deceptive}.
It was formalized as a path planning problem by Masters and
Sardina~\cite{masters2017deceptive}, who introduced the objective of
minimizing early goal recognition and defined the notion of a last
deceptive point.
Building on a long-standing deception taxonomy of simulation (showing the false) and dissimulation (hiding the real), several works have converged on two practical motion strategies: exaggeration and ambiguity~\cite{dragan2015deceptive,savas2021}. Beyond generating deceptive trajectories, {researchers also study} (i) observer/goal-recognition models and prediction priors, including max-entropy and resource-allocation variants~\cite{chen2022}, (ii) algorithms ranging from trajectory optimization to learning-based and RL pipelines~\cite{fatemi2024dpp}, and (iii) evaluation protocols and metrics, including human-subject studies~\cite{dragan2015deceptive,gutierrez2025dppp}.

Optimization-based DPP methods  typically solve from a start position to a final goal
and produce a single deceptive path plan, largely in single-agent and static environment  settings~\cite{chen2022,savas2021,xu2020single,fu2022almost,lv2024optimal,price2023domain,lenhard2025cadetopath,xu2020scalable}. 
Works that admit environmental changes typically re-solve end-to-end starting from the current state when conditions shift~\cite{fatemi2024dpp,xue2025efficient,chen2024dpp}.

The foundational work of Savas et al.  \cite{savas2021} developed an optimization framework resulting in a linear program (LP) formulation for balancing an optimal (e.g., shortest) path to the true goal with deceptive deviation with respect to false goals, and incorporating probabilistic constraint satisfaction.
Agents' behavior is modeled as a Markov decision process, and entropy is used to model an observer's prediction of the path. Human studies validate the deception. Our work is inspired by the LP planner, providing a practical local planner using similar costs and enabling dynamic adaptation. We experimentally compare with the LP-based planner, showing that it is possible to employ our receding horizon method and achieve a globally optimal path.

Methods for multi-agent DPP are limited. An early heuristic search method uses constraints and lays out waypoints~\cite{root2005,root2005deceptive}. Extensions to \cite{savas2021} include a form of multi-agent deception cast in an
MDP framework with the goal of achieving a desired final distribution of agents over goal
states \cite{chen2022}. A two-phase solution is developed, switching from a deception phase with a 
false goal state distribution to a goal-directed phase ending in the true goal distribution.  
Other extensions to \cite{savas2021} include deceiving a supervisor by minimizing deviation from the expected nominal policy so that at least one agent will achieve a hidden alternate objective \cite{probine2024}, and altering the path based on an adversarial intervention model \cite{suttle2025}.

Learning-based path planning methods face the challenge of generalization over environments, goal and false-goal placement, and forms of deception. In \cite{fatemi2024dpp}, a GNN-based inference engine was developed over a local graph of possible moves from the agent's current location. This used the LP solution as a training signal and was shown to generalize over different graph topologies and deception weights. The local graph can be obtained by spatial sampling around the agent, providing a link to continuous action and state space control.

We emphasize that our framework is not learning-based and does not incorporate training. Note that the Boltzmann distribution has been classically used in general reinforcement learning (RL) and bandit problems to balance exploration and exploitation. However, relative to classical RL, which manages exploration–exploitation via learned value/policy estimates and explicit exploration schemes (e.g., $\varepsilon$-greedy, UCB, and Thompson sampling)~\cite{sutton2018reinforcement,auer2002finite,bubeck2012regret,thompson1933}, our planner induces stochasticity directly at planning time without any training phase.
Thus, while our method is similar to softmax action selection used in RL, we employ tunable deterministic costs that incorporate deception and constraints and map these to a random policy through normalization of the Boltzmann distribution.

Various comparison metrics in the above references 
are used 
to quantify the deviation from an expected policy; see also~\cite{gutierrez2025dppp}. Our framework follows this arc but emphasizes a simple, on-the-fly stochastic policy, derived from the Boltzmann distribution over short-horizon action-sequence candidates.

\section{Single-Agent Method}\label{AL}

In this section, we develop a Boltzmann-based formulation for constructing
single-agent DPP policies. The Boltzmann distribution originates from statistical mechanics, where the probability of a microscopic state $x$ with energy $E(x)$ is proportional to $\exp\!\left(-\frac{E(x)}{\theta_{\mathrm{temp}}\, C_{\mathrm{B}}}\right)$. The temperature parameter $\theta_{\mathrm{temp}}$ scales the exponent and thereby controls the spread of probability mass, and $C_{\mathrm{B}}$ denotes the Boltzmann constant \cite{boltzmann1877}. By analogy, the same exponential form is used to map scalar costs over candidate action sequences to a probability distribution. 
Let $A_t$ denote a finite candidate path or action sequence 
starting at time $t$, and let $C(A_t) \in \mathbb{R}$ denote its associated deterministic scalar cost.
The resulting Boltzmann policy is based on the probability distribution over candidate action sequences given by
\setlength{\abovedisplayskip}{4pt plus 0pt minus 1pt}%
\setlength{\belowdisplayskip}{4pt plus 0pt minus 1pt}%
\setlength{\abovedisplayshortskip}{2pt plus 0pt minus 1pt}%
\setlength{\belowdisplayshortskip}{2pt plus 0pt minus 1pt}%
\begin{equation}
\Phi_t(A_t) = 
\frac{e^{-\lambda C(A_t)}}{\sum_{\tilde A_t \in \mathcal{A}_t} e^{-\lambda C(\tilde A_t)}} ,
\label{eq:boltzmann_policy_1}
\end{equation}
where {scalar} $\lambda > 0$ is the rationality parameter, playing a role analogous
to the inverse of the temperature scaling factor $(\theta_{\mathrm{temp}}\, C_{\mathrm{B}})^{-1}$. 
Note that $\lambda$ tunes the level of randomness independently of the deterministic cost, and smaller $\lambda$ yields more randomness.
The denominator normalizes the ratio to always yield a valid probability, summing over all possible sequences in set $\mathcal{A}_t$ using dummy variable $\tilde A_t$. 

We refer to $C(A_t)$ as cost, and under (\ref{eq:boltzmann_policy_1}), candidate action sequences with smaller costs are assigned higher probabilities. 
We may also employ incentives that lower the cost, i.e., that move the cost to the left on the real axis, and subsequently yield a higher probability through (\ref{eq:boltzmann_policy_1}). 


{
This construction yields a stochastic policy that samples action sequences according to their relative costs, rather than deterministically selecting a single optimal action sequence. 
The exact form of $C(\cdot)$ is user- and scenario-dependent and will typically have two or more terms in a weighted sum to incorporate deception and constraints, as described below.
The Boltzmann-based approach naturally extends to multi-agent path planning by including joint cost terms in $C(\cdot)$ and sampling from the resulting joint distribution;
we develop this in Section \ref{MA}.
}

Notation for the system variables is given at the top of Table \ref{tab:notation_merged}, and tunable parameters for the single-agent case are listed in Table~\ref{tab:parameters}. 
To facilitate the algorithmic description, we consolidate the system variables into the set $\Gamma_t$ defined as
{
\setlength{\abovedisplayskip}{2pt plus 0pt minus 1pt}%
\setlength{\belowdisplayskip}{2pt plus 0pt minus 1pt}%
\setlength{\abovedisplayshortskip}{2pt plus 0pt minus 1pt}%
\setlength{\belowdisplayshortskip}{4pt plus 0pt minus 1pt}%
\begin{equation}
\Gamma_t = \left\{\, s_0,\ s_t,\ \hat{s}^j_{t+K_t},\ A^j_t,\ a^{jm}_t,\ \mathcal{A}_t,\ A^*_t,\ a_t^{*m} \,\right\}
\label{eq:gamma_vector}
\end{equation}
}
and the agent-tunable
parameters into the set $\Theta_t$, given by 
\begin{equation}
\Theta_t =
\left\{
\begin{array}{l}
G_t,\ F_t^{1},\ \dots,\ F_t^{z},\ \lambda_t,\ B_t,\ K_t, \\
M_t,\ w_t^{1},\ \dots,\ w_t^{4},\ \delta_t^{3},\ \delta_t^{4}
\end{array}
\right\} .
\label{eq:theta_vector}
\end{equation}

The agent’s initial state is denoted $s_0$, its current state at time~$t$ by~$s_t$, and its true goal by~$G_t$ (or simply $G$ if not time-varying).  
We let $\tau$ denote the planned state trajectory induced by executing an action sequence ${A}_t$.
The agent starts at $s_0$, knows its current state~$s_t$ and true goal~$G$, and moves one unit of distance per time step. The agent can adjust its parameters in $\Theta_t$ and stops upon reaching its goal state $G$.

Next, we describe the generation of candidate action sequences.
Let $\mathcal{G}_{\mathcal{A}}(\cdot,\cdot)$
denote a generator of the feasible set $\mathcal{A}_t$, starting from state $s_t$ for length $K_t$, so \mbox{$\mathcal{A}_t = \mathcal{G}_{\mathcal{A}}(s_t, K_t)$} indicates generation of the entire set.
The $j$-th candidate action sequence is $A^j_t = [a^{j1}_t, \dots, a^{jK_t}_t] \in \mathcal{A}_t$, and we denote its resulting end state as $\hat{s}_{t+K_t}^j$. 
The end state is obtained via
$\hat{s}^j_{t+K_t} = \mathcal{T}_{K_t}(s_t, A^j_t),$
where $\mathcal{T}_{K_t}(\cdot,\cdot)$ generically denotes the $K_t$-step transition to the end state for the particular candidate action sequence $A^j_t$ starting from $s_t$.

\begin{table}[H]
\caption{System variables. Superscript $i$ denotes the $i$-th agent. For the single-agent case, the superscript is omitted, e.g., $s_t$ corresponds to $s^i(t)$. The index $m \in \{1,\dots,K_t\}$ denotes the step within a $K_t$-step sequence.} 
\label{tab:notation_merged}

\begin{tabularx}{\textwidth}{lL}
\toprule
\textbf{Symbol} & \textbf{Description} \\
\midrule
$t$ & Current time step \\
$s_0$ & Initial state (start location)  \\
$s_t$ & State at time $t$ \\
$a_t^{*m}$ & $m$-th action in the selected plan $A_t^*$ \\
$\mathcal{A}_t$ & Feasible set of $K_t$-step sequences at time $t$ \\
$A_t^*$ & Selected action sequence at time $t$ \\
$A_t^j$ & $j^{th}$ candidate action sequence at time $t$ of length $K_t$ \\
$a_t^{jm}$ & Action at step $m$ within sequence $A_t^j$ \\
$\hat{s}^j_{t+K_t}$ & End state for sequence $A_t^j$ \\
$\tau$ & Planned trajectory \\
\midrule
& {\bf Multi-Agent Case} \\
\midrule
$s_0^i$ & Initial state (start location) of agent $i$ \\
$N$ & Total number of agents \\
$s^i(t)$ & State of agent $i$ at time $t$ \\
$S(t)$ & Joint state of all agents at time $t$ \\
$\hat{s}^{i,j}(t + K^i(t))$ & End state of agent $i$ under $A^{i,j}(t)$ \\
$A^{i,j}(t)$ & $j^{th}$ candidate action sequence for agent $i$ at time $t$
 \\
$\mathcal{A}^i(t)$ & Set of feasible candidate action sequences for agent $i$ \\
$\mathcal{A}(t)$ & Joint set of feasible candidate sequences for all agents \\

\midrule

$\mathbf{A}(t)$ & Joint candidate action sequence at time $t$ (one sequence per agent) \\
$\hat{S}(t)$ & Collection of all end states under $\mathbf{A}(t)$ \\
$\mathbf{C}_{\mathcal{A}}(t)$ & Cost matrix \\
$\mathbf{A}^*(t)$ & Selected multi-agent action plan \\
\bottomrule
\end{tabularx}

\end{table}
\vspace{-12pt}

\begin{table}[H]

  \caption{Tunable single-agent parameters.}
  \label{tab:parameters}
  \renewcommand{\arraystretch}{1.2}
  \begin{tabularx}{\textwidth}{lL}
    \toprule
    \textbf{Parameter} & \textbf{Description} \\ 
    \hline
    $G_t$ & True goal at time $t$ \\
    $F_t^z$ & False goals at time $t$, $z \in \{1, 2, \dots\}$ \\
    $\lambda_t$ & Rationality parameter \\
    \hline
    $K_t$ & {{Planning trajectory length} }\\
    $M_t$ & {Number of steps before replanning, $M_t \leq K_t$\ }\\
    \hline

    $w^1_t, \dots, w^4_t$ & Cost term coefficients \\
    $B_t$ & Optional time budget constraint \\
    $\delta^3_t, \delta^4_t$ & Binary flags for optional cost terms \\
    \bottomrule
  \end{tabularx}
\end{table}

{We define the overall cost function 
as a weighted sum of scalar terms, }
computed at time $t$ based on the current state $s_t$ and the trajectory with end state $\hat{s}_{t+K_t}^j$,
given by
\setlength{\abovedisplayskip}{2pt plus 0pt minus 1pt}%
\setlength{\belowdisplayskip}{2pt plus 0pt minus 1pt}%
\setlength{\abovedisplayshortskip}{2pt plus 0pt minus 1pt}%
\setlength{\belowdisplayshortskip}{2pt plus 0pt minus 1pt}%
\begin{equation}
\begin{split}
\mathcal{C}(\Theta_t, \Gamma_t) =\ & w^1_t C_{\mathrm{goal}}(G_t, s_t, \hat{s}^j_{t+K_t}) \\
&+ w^2_t C_{\mathrm{decep}}(G_t, \{F_t^z\}z, s_t, \hat{s}^j_{t+K_t}) \\
&+ \delta^3_t w^3_t C_{\mathrm{time}}(t, B_t, s_t, \hat{s}^j_{t+K_t}) \\
&+ \delta^4_t w^4_t C_{\mathrm{smooth}}(s_t, \hat{s}^j_{t+K_t}).
\end{split}
\label{eq:cost_function}
\end{equation}

{Each cost term in (\ref{eq:cost_function}) 
returns a 
scalar in $\mathbb{R}$.}  $C_{\mathrm{goal}}$ encourages reaching $G_t$, and $C_{\mathrm{decep}}$ encodes the deception objective 
relative to false goals 
$\{F_t^z\}z$. 
Optional costs include $C_{\mathrm{time}}$, which enforces a time-to-goal constraint specified by $B_t$, and $C_{\mathrm{smooth}}$, which penalizes non-smooth motion. {These will be described further in the examples presented in Section \ref{ER}. This cost structure is flexible and can be varied based on user-defined objectives, incentives, and scenario-specific requirements.}

By substituting $\mathcal{C}(\Theta_t, \Gamma_t)$ from \eqref{eq:cost_function} into~\eqref{eq:boltzmann_policy_1}, the probability of action sequence $A_t$ is given by
\setlength{\abovedisplayskip}{2pt plus 0pt minus 1pt}%
\setlength{\belowdisplayskip}{2pt plus 0pt minus 1pt}%
\setlength{\abovedisplayshortskip}{2pt plus 0pt minus 1pt}%
\setlength{\belowdisplayshortskip}{2pt plus 0pt minus 1pt}%
\begin{equation} 
\Phi_t(A_t \mid \Theta_t) = \frac{e^{-\lambda_t \mathcal{C}(\Theta_t, \Gamma_t)}}{\sum_{\tilde{A} \in \mathcal{A}_t} e^{-\lambda_t \mathcal{C}(\Theta_t, \tilde{\Gamma}_t)}} .
\label{eq:boltzmann_policy_updated} 
\end{equation} 
Here, the denominator sum is over all possible action sequences $\tilde{A} \in \mathcal{A}_t$, normalizing the ratio to provide a legitimate probability.
The agent samples this distribution, represented as
\begin{equation}
A^*_t \sim \Phi_t(A_t \mid \Theta_t), 
\end{equation}
to obtain a $K_t$-step action sequence $A^*_t = ( a_t^{*1}, \dots, a_t^{*K_t} )$.  
The agent can then follow the selected short-distance trajectory $\tau$ that ends in state $s_{t+1}$. The agent's parameters and cost function are then updated if desired, and the planning process is repeated. 

The single-agent deceptive path planner is summarized in Algorithm~\ref{alg:single_agent_dpp}, with the inputs being parameter set $\Theta_t$ and starting state $s_t$, and the output being the planned trajectory $\tau$. 
The algorithm statement loops over successive short-distance plans until the goal $G_t$ is reached (the while-loop at line 2) and outputs a longer concatenated trajectory. The algorithm is easily modified to provide only one, or more, successive plans as desired. 
Line 3 generates the feasible candidate set of action sequences. Lines 4--7 compute the end state and cost for each candidate action sequence $A_t^j \in \mathcal{A}_t$. 
Line 8 generates the distribution of candidate sequences, which is sampled in line 9 to yield the selected sequence $A_t^*$.
The loop in lines 10--17 allows for executing only the first $M_t \leq K_t$ steps if desired. The end state is recorded in line 11, and the trajectory is appended to a longer stored trajectory in line 12.  
The loop at line 13 terminates the algorithm if the goal $G_t$ is reached. 
Finally, the trajectory $\tau$ is returned.

\begin{algorithm}[H]
\caption{Single-Agent Deceptive Path Planner} 
\label{alg:single_agent_dpp}
\begin{algorithmic}[1] 
\Require $\Theta_t, s_t$ 
\State $\tau \gets [s_t], t \gets 0$ \hfill $\triangleright$ Initialize trajectory
\While{$s_t \neq G_t$} \hfill $\triangleright$ Loop until goal is reached
    \State $\mathcal{A}_t \gets \mathcal{G}_{\mathcal{A}}(s_t, K_t)$ \hfill $\triangleright$ Generate feasible set
    \For{each $A^j_t \in \mathcal{A}_t$} 
        \State $\hat{s}^j_{t+K_t} \gets \mathcal{T}_{K_t}(s_t, A^j_t)$ \hfill $\triangleright$ Compute $K_t$ step final state
        \State $c_j \gets \mathcal{C}(\Theta_t, \Gamma_t^j)$ \hfill $\triangleright$ Compute cost
    \EndFor
    \State $\Phi_t \gets \Phi_t(A_t \mid \Theta_t)$ \hfill $\triangleright$ Construct policy per Equation \eqref{eq:boltzmann_policy_updated}
    \State $A_t^* \sim \Phi_t$ \hfill $\triangleright$ Sample the policy
    \For{$m \in \{1, \dots, M_t\}$} \hfill $\triangleright$ Execute $M_t \leq K_t$ steps
        \State $s_{t+1} \gets T(s_t, a_t^{*m})$ \hfill $\triangleright$ Advance to end state
        \State $\tau \gets \tau \cup \{s_{t+1}\}$  \hfill $\triangleright$ Append to full trajectory
        \If{$s_{t+1} = G_t$} \hfill $\triangleright$ Check arrived at true goal
            \State \textbf{break} 
        \EndIf
        \State $t \gets t + 1$ \hfill $\triangleright$ Increment global clock
    \EndFor
\EndWhile
\State \textbf{return} $\tau$ \hfill $\triangleright$ Return end-to-end planned trajectory 
\end{algorithmic}
\end{algorithm}

\section{Multi-Agent Method}\label{MA}

Consider $N$ agents whose
joint state at time~$t$ is  
$S(t) = (s^1(t), \dots, s^N(t))$, and 
let the collection of agent scalar rationality parameters be denoted by \( \{{\lambda}^i(t)\}_{i=1}^N \).  
The true goal of agent~$i$ at time~$t$ is denoted $G^i(t)$, or $G^i$ when the goal is time-invariant. 
Similar to the single-agent case, the 
system variables for the multi-agent algorithm are listed in Table~\ref{tab:notation_merged}, and the agents' tunable parameters are listed in Table~\ref{parameter_multi} (generalizing Table~\ref{tab:parameters}).

\begin{table}[H]

\caption{Tunable $i$-th agent parameters (multi-agent case).}
\label{parameter_multi}
\begin{tabularx}{\textwidth}{lL}
\toprule
\textbf{Symbol} & \textbf{Description} \\
\midrule

\( G^i(t) \) & True goal of agent \( i \) at time \( t \) \\

$\{F^i_z(t)\}_{z=1}^{n(i)}$ & Set of $n(i)$ false goals for agent $i$ at time $t$ \\

\( K^i(t) \) & Number of planning-horizon steps for agent \( i \) \\

\( M^i(t) \) & Number of steps executed before replanning \\

\( B^i(t) \) & Max total allowed (time) steps for agent \( i \) \\

\( \{w^i_\ell(t)\}_{\ell=1}^{4} \) & Weights for $i$-th agent cost terms  \\

\( \{{\lambda}^i(t)\}_{i=1}^N \) & Rationality parameters for each agent \\

\({ \lambda}(t) \) & Rationality vector \\

\bottomrule
\end{tabularx}
\end{table}

As we did in the single-agent case, for ease of notation we define sets of parameters as follows.
The set of tunable parameters for all $N$ agents at time $t$ is given by
\begin{equation}
\boldsymbol{\Omega}(t) = 
\left\{ \boldsymbol{\Omega}^i(t) \right\}_{i=1}^N
\label{eq:omega_vector}
\end{equation}
where the subset of parameters for agent $i$ is
\begin{equation}
\boldsymbol{\Omega}^i(t)
=
\left\{G^i(t), \{F^i_z(t)\}_{z=1}^{n(i)}, K^i(t), M^i(t), B^i(t), \{w^i_\ell(t)\}_{\ell=1}^{4}, \lambda^i(t)\right\}.
\label{eq:omega_i_detail}
\end{equation}
Similarly, the set of all system variables is 
\begin{equation}
\begin{aligned}
\boldsymbol{\Upsilon}(t)= & \left\{ 
\left(s^i_0,s^i(t),A^{i,j}(t),\hat{s}^{i,j}\!\left(t+K^i(t)\right) \right)_{i=1}^{N}, 
\left(\mathcal{A}^i(t)\right)_{i=1}^{N},N,S(t), \right. \\
& \left. \mathcal{A}(t),\hat{S}(t),\mathbf{A}(t),
\mathbf{C}_{\mathcal{A}}(t),\mathbf{A}^*(t) \right\},
\end{aligned}
\label{eq:upsilon_vector}
\end{equation}
and we denote $\boldsymbol{\Upsilon}^i(t)$ 
as the set of system parameters for the $i$-th agent, drawn from $\boldsymbol{\Upsilon}(t)$.

We assume each agent knows the joint state $S(t)$, the goals and false goals for all agents, and shares collective knowledge of the variables and parameters  $\boldsymbol{\Omega}(t)$ and $\boldsymbol{\Upsilon}(t)$. These assumptions can be relaxed, for example, to form and disband subsets of coupled agents. 
To illustrate the planning in our grid-world simulations, each agent moves one unit distance per time step, and agents stop when their goal state is achieved.

{
Next, we describe the generation and cost assignment for candidate joint trajectory sequences, leading to the Boltzmann-based policy form in the multi-agent case.
Let $A^{i,j}(t) = [a^{i,j}_1(t), \dots,a^{i,j}_{K_i(t)}(t)]$ denote a feasible action sequence for the $i$-th agent over $K^i(t)$ steps.
Beginning in the current state 
$s^i(t)$ and executing action sequence 
$A^{i,j}(t)$, this $j$-th trajectory ends in state $\hat{s}^{i,j}(t + K^i(t))$. Now, let $\mathcal{A}^i(t)$ be the collection of all feasible action sequences for the $i$-th agent beginning at state $s_i(t)$ at time $t$. Thus, $\mathcal{A}^i(t)$ is a set, and 
$A^{i,j}(t)  \in \mathcal{A}^i(t)$. 
}

{
A joint candidate action is denoted
\begin{equation}
\mathbf{A}(t)
=
\big(A^{1,j_1}(t),\ldots,A^{N,j_N}(t)\big),
\label{eq:joint_action_definition}
\end{equation}
and contains one candidate action sequence for each of the $N$ agents; $j_i$ is the index of the candidate action sequence selected for agent $i$. 
Each $A^{i,j_i}(t)$ has dimensions $1 \times K^i(t)$, with zero padding if some agent(s) have a shorter planning horizon than others. Therefore, $\mathbf{A}(t)$ consists of $N$ rows, where the $i$-th row has dimension $1 \times \text{max}_i K^i(t)$. 
}

For a given joint action candidate $\mathbf{A}(t)$, we collect the ending state for each agent. The $i$-th agent begins in state $s_i(t)$ and follows $A^{i,j_i}(t)$, arriving at end state $\hat{s}^{i,j_i}(t + K^i(t)).$ The collection of end states for the candidate joint action is denoted
\begin{equation}
\hat{S}(t) = 
\big\{ \hat{s}^{i,j_i}(t + K^i(t)) \big\}_{i=1}^N.
\label{eq:predicted_joint_state}
\end{equation}

{Next, we evaluate the cost for each joint action candidate 
$\mathbf{A}(t)$. The total cost consists of two parts: (i) costs associated with the individual $i$-th agent and its candidate trajectory, and (ii) joint interaction costs that couple agents. 
Depending on the specific cost definitions, these may be evaluated step-by-step along a trajectory, or based only on the end states $\hat{S}(t)$.}

We specify the costs as follows. First, similar to the single-agent case, we evaluate the cost of trajectory sequence $A^{i,j}(t)$ for the $i$-th agent. This depends on the $i$-th agent's parameters $\boldsymbol{\Omega}^i(t)$ and $\boldsymbol{\Upsilon}^i(t)$, the state transitions and final state for a given action sequence $A^{i,j}(t)$, and the specific form of the individual cost terms that are summed.  These per-agent scalar cost components are given by
\begin{equation}
\mathbf{c}^{i,j}(t) = 
\begin{bmatrix}
C^i_{\mathrm{goal}}(t) \\
C^i_{\mathrm{decep}}(t) \\
C^i_{\mathrm{time}}(t) \\
C^i_{\mathrm{smooth}}(t)
\end{bmatrix} .
\label{eq:individual_cost_function}
\end{equation}
In the deception context of this paper, we always include $C^i_{\mathrm{goal}}(t)$ and $C^i_{\mathrm{decep}}(t).$  The costs in (\ref{eq:individual_cost_function}) 
reflect those used in the single-agent case, as in Equation (\ref{eq:cost_function}), including the optional costs $C^i_{\mathrm{time}}(t)$ and $C^i_{\mathrm{smooth}}(t)$, which we use in some of our examples in Section \ref{ER}.

Second, for the multi-agent case, we introduce $L$ scalar interaction costs, denoted as
\begin{equation}
\boldsymbol{\phi}(t)
=
\begin{bmatrix}
C_{(1)}(t)\\
\vdots\\
C_{(L)}(t)
\end{bmatrix}.
\label{eq:interaction_cost_vector}
\end{equation}
These costs reflect coupling among agents; see the examples in Section \ref{ER}.

{
Now, we combine the above costs in preparation for the joint Boltzmann formulation. For any joint candidate action $\mathbf{A}(t)$, we define the full cost vector as
\begin{equation}
\mathbf{C}_{\mathbf{A}}(t) =
\left[
C^{1,j_1}(t),\,
\dots,\,
C^{N,j_N}(t),\,
C^{\mathrm{int}}(t)
\right]^{\top}.
\label{eq:full_cost_vector}
\end{equation}
The $N$ leading terms in $\mathbf{C}_{\mathbf{A}}(t)$ are the costs for each individual agent, obtained from the weighted sum of costs in (\ref{eq:individual_cost_function}), similar to Equation (\ref{eq:cost_function}). Recall that in (\ref{eq:cost_function}), the combination weights are denoted $\omega_t$, and the $\delta_t$ are binary in $\{0,1\}$ to optionally include cost terms. 
The last term in (\ref{eq:full_cost_vector}) is the sum of interaction costs given by 
\begin{equation}
C^{\mathrm{int}}(t) = \sum_{l=1}^L C_{(l)}(t). 
\end{equation}
}

Generalizing the single-agent case, we apply the Boltzmann approach for $N$ agents. The probability of selecting a specific joint action candidate
$\mathbf{A}(t)$ is 
\begin{equation}
\eta_{\mathbf{A}}(t)
=
\frac{
e^{-\lambda(t)^\top \mathbf{C}_{\mathbf{A}}(t)}
}{
\displaystyle
\sum_{\widetilde{\mathbf{A}}(t)\in\mathcal{A}(t)}
e^{-\lambda(t)^\top
\mathbf{C}_{\widetilde{\mathbf{A}}}(t)}
}. 
\label{eq:boltzmann_distribution_compute}
\end{equation}
We introduced the rationality vector defined by
\begin{equation}
\lambda(t) = [\lambda^1(t), \cdots, \lambda^N(t), \lambda^{int}]^{\top}, 
\end{equation}
which collects the rationality parameters of each agent and an additional parameter $\lambda^{int}$ that weights $C^{\mathrm{int}}(t),$ so that $\lambda(t)\in\mathbb{R}^{N+1}$ matches the dimension of $\mathbf{C}_{\mathbf{A}}(t).$ 
Here, 
$\widetilde{\mathbf{A}}(t)$ denotes a joint action candidate from set $\mathcal{A}(t)$, 
and the normalizing sum in the denominator is over all joint action candidates to ensure $\eta_{\mathbf{A}}(t)$ is a valid probability. 

Finally, the joint action $\mathbf{A}^*(t)$ is selected by sampling from $\mathcal{A}(t)$ according to the distribution
$\boldsymbol{\eta}_\mathcal{A}(t)$, represented by
\begin{equation}
\mathbf{A}^*(t) \sim \boldsymbol{\eta}_\mathcal{A}(t) ,
\label{eq:action_sampling}
\end{equation}
so that $\mathbf{A}^*(t)$ contains the planned trajectories for each agent.

\textls[-15]{The multi-agent DPP method is summarized in
Algorithm~\ref{alg:multi_agent_dpp}. The inputs are $\boldsymbol{\Omega}(t)$ (multi-agent planning parameters) and the joint state $S(t)$;
the output is the joint action sequence $\mathbf{A}^*(t)$. Lines 2--4 generate feasible candidate action sequences
for each agent. Lines 5--8 find the end state for each feasible agent sequence and compute the corresponding per-agent costs. 
The loop in lines 9--15 computes the probability for each feasible joint action sequence. Line 10 collects the end states.  Lines 11--12 compute and aggregate the interaction costs. Line 13 constructs the cost vector, and line 14 computes the probability of the joint action sequence.  Finally, line 16 samples the joint planning policy, and the joint plan is returned in line 17. }

\begin{algorithm}[H]
\caption{Multi-Agent Deceptive Path Planner}
\label{alg:multi_agent_dpp}
\begin{algorithmic}[1]

\State \textbf{Input:} $\boldsymbol{\Omega}(t)$,
$S(t)=(s^1(t),\dots,s^N(t))$

\ForAll{agents $i$}
    \State
    $\mathcal{A}^i(t)
    \gets
    \bigl[s^i(t),K^i(t)\bigr]$
    \Comment{Generate feasible action sequences}
\EndFor

\ForAll{$i$ and $A^{i,j}(t)\in\mathcal{A}^i(t)$}

    \State
    $\hat{s}^{i,j}(t+K^i(t))
    \gets
    \bigl[s^i(t),A^{i,j}(t)\bigr]$
    \Comment{Compute end state}

    \State
    $\mathbf{c}^{i,j}(t)
    \gets
    \begin{bmatrix}
    C^i_{\mathrm{goal}}(t)\\
    C^i_{\mathrm{decep}}(t)\\
    C^i_{\mathrm{time}}(t)\\
    C^i_{\mathrm{smooth}}(t)
    \end{bmatrix}$
    \Comment{Compute per-agent costs}

\EndFor

\ForAll{$\mathbf{A}(t)$}

    \State
    $\hat{S}(t)
    \gets
    \left\{
    \hat{s}^{i,j_i}(t+K^i(t))
    \right\}_{i=1}^{N}$
    \Comment{Collect end states}

    \State
    $\boldsymbol{\phi}(t)
    \gets
    \begin{bmatrix}
    C^{(1)}(t)\\
    \vdots\\
    C^{(L)}(t)
    \end{bmatrix}$
    \Comment{Compute interaction costs}

    \State
    $C_{\mathrm{int}}(t)
    \gets
    \displaystyle\sum_{\ell=1}^{L} C^{(\ell)}(t)$
    \Comment{Aggregate interaction costs}

    \State
    $\mathbf{C}_{\mathbf{A}}(t)
    \gets
    \left[
    C^{1,j_1}(t),\dots,
    C^{N,j_N}(t),
    C_{\mathrm{int}}(t)
    \right]^{\top}$
    \Comment{Construct cost vector}

    \State
    $\eta_{\mathbf{A}(t)}
    \gets
    \frac{
    e^{-\boldsymbol{\lambda}(t)^{\top}
    \mathbf{C}_{\mathbf{A}}(t)}
    }{
    \displaystyle
    \sum_{\tilde{\mathbf{A}}(t)\in\mathcal{A}(t)}
    e^{-\boldsymbol{\lambda}(t)^{\top}
    \mathbf{C}_{\tilde{\mathbf{A}}}(t)}
    }$
    \Comment{Compute Boltzmann probability}

\EndFor

\State
$\mathbf{A}^*(t)
\sim
\eta_{\mathbf{A}(t)}$
\Comment{Sample the joint policy}

\State \Return $\mathbf{A}^*(t)$
\Comment{Return joint plan}

\end{algorithmic}
\end{algorithm}

Algorithm~\ref{alg:multi_agent_dpp} may be executed repeatedly, beginning with the current joint state and any changes to the parameters in $\boldsymbol{\Omega}(t)$. The planner ends when agents reach their goal states.

\subsection{Information Sharing Among Agents}

Multi-agent path planning is a classic problem with combinatorial complexity, e.g., see  \cite[Sect.~7.2]{LavallePlanningBook}. 
In our multi-agent setting, only a small amount of information is assumed available for coordinated planning when repeatedly applying Algorithm~\ref{alg:multi_agent_dpp}. 
Once initialized, only changes to the tunable parameters in $\boldsymbol{\Omega}(t)$ need to be shared. This includes parameters affecting individual agent costs, interaction costs, and constraints. As we show in the examples, a time schedule may be set to change some agents' parameters, where agents otherwise proceed independently or with dependent sub-groups. Algorithm~\ref{alg:multi_agent_dpp} also requires the initial joint state, i.e., the end state for each agent after executing the most recent plan.

Our scheme yields joint rollouts from random sampling, as in  Equation (\ref{eq:action_sampling}) and line 16 in  Algorithm~\ref{alg:multi_agent_dpp}. 
If it is assumed the agents will reach their end states with high probability, then even this need not be communicated for every planning round.  
A further simplification is to employ independent rollouts, where each agent evaluates its own candidate sequences $A^{i,j}(t)$ and shares only its final state $\hat{s}^{i,j}(t+K^i(t))$.  
Agents can also collaborate without evaluating all
joint profiles by intermittently sharing summaries of their local
policies, exchanging only the parameters that influence coupled costs, or interleaving local Boltzmann updates with occasional global synchronization. These quasi-independent strategies aim to preserve coordination while avoiding the communications and computational burden of full joint updates.
This provides design flexibility for collaboration and
rollouts, and the potential for decentralization. 

The coupled planning problem is often cast in robotics, where moving obstacles and robot collisions must be avoided. Many heuristics have been studied with sophisticated implementations in complex environments, e.g., see \cite{Prorok2026}.
We believe that our framework has sufficient generality to enable incorporation into such existing planning pipelines.

\subsection{Complexity and Scaling}\label{CA}

{
To study complexity and scaling, we consider the worst case where all feasible action sequences are enumerated. 
Let $B$ denote the branching factor, i.e., the number of feasible actions from a state. 
For a single agent at each receding-horizon update, the planner enumerates all feasible action sequences of length $K$. Each planning step has a branching factor $B$, so the worst-case candidate set contains $\mathcal{O}(B^{K})$ sequences.
Each candidate requires a $K$-step rollout and constant-time cost evaluation, giving overall complexity
$\mathcal{O}(B^{K} K)$.
}

Next, we consider the multi-agent case.
As above, let each agent have a branching factor $B$ and a $K$-step lookahead, producing a single-agent local candidate set of size $\mathcal{O}(B^{K})$. 
If the team cost function decomposes (separable cost case), then 
each agent performs an independent update with cost $\mathcal{O}(B^{K}K)$, yielding a total complexity of $\mathcal{O}(N B^{K} K)$, i.e., the overall cost scales linearly with the number of agents $N.$

{When the team cost is not separable, then evaluating a Boltzmann policy requires considering all feasible joint candidate action sequences whose set size is}
\begin{equation} 
|\mathcal{A}(t)| = \prod_{i=1}^N |\mathcal{A}^i(t)| \approx \mathcal{O}(B^{KN}) ,
\end{equation}
where $|\cdot|$ denotes set cardinality.
Each candidate joint action requires a $K$-step rollout and cost evaluation, giving an overall complexity of
$\mathcal{O}(B^{KN} K)$,
which grows exponentially with $N$ under full coupling.

When $K$ is relatively small, the update costs remain relatively lightweight. 
In practice, the complexity may be much smaller; the planner does not need to enumerate all $B^{K}$ branches in the single-agent case, or explore the full $B^{KN}$ sized space for the multi-agent case. 
Standard search techniques (e.g., pruning, heuristic rollouts, or beam-search–style truncation) can limit the number of explored sequences, ensuring that only a small subset of high-value branches is evaluated at each step.  
Thus, the empirical complexity per update can be substantially below the worst-case bound for receding horizon planning.

\section{Experimental Results}\label{ER}

In this section, we present a variety of experiments, showing 
how single and multiple agents can maintain deception while incorporating 
local and resource constraints. 
We demonstrate dynamic deception induced through time-varying parameters, and we also show that the algorithm can reproduce optimal end-to-end paths using updated local planning by using the LP method from \cite{savas2021} as a baseline.
The experiments illustrate the flexibility and dynamic adaptation of the framework, revealing the potential for use in many different settings.

\subsection{Single-Agent DPP: Exaggeration and Ambiguity}

\subsubsection{Exaggeration}
\label{exageeration}
An exaggerated path seemingly targets a false goal while eventually arriving at the true goal. This represents a tradeoff between an efficient path to the true goal and deviation towards a false goal, e.g., see the discussion in~\cite{savas2021}.
We instantiate the planner in Algorithm~\ref{alg:single_agent_dpp} on a $375{\times}375$ grid (four-neighbor moves), with a small margin from the boundary. We perform $500$ Monte Carlo trials and accumulate visit-frequency heatmaps.

We use Euclidean distances to true and false goals given by
\begin{equation}
\Delta_{\mathrm{true}}(s_t,s_{t+1})
= \big[\, d(s_t,G) - d(s_{t+1},G) \,\big]_+ 
\end{equation}
\begin{equation}
\Delta_{\mathrm{false}}(s_t,s_{t+1})
= \big[\, d(s_t,F) - d(s_{t+1},F) \,\big]_+ 
\end{equation}
where $[x]_+ \!=\!\max\{x,0\}$, $G$ is the true goal, and $F$ is the false goal.
We adopt the cost in \eqref{eq:cost_function} with $K\!=\!1$ and deactivate time and smoothness terms
($\delta_3\!=\!\delta_4\!=\!0$). The per-step cost components are now
\begin{equation}
C_{\mathrm{goal}}(G,s_t,s_{t+1}) \triangleq -\,\Delta_{\mathrm{true}}(s_t,s_{t+1}) 
\label{eq:components_goal}
\end{equation}
\begin{equation}
C_{\mathrm{decep}}(G,F,s_t,s_{t+1}) \triangleq -\,\Delta_{\mathrm{false}}(s_t,s_{t+1}) 
\label{eq:components_decep}
\end{equation}
so moving closer to a target reduces cost. 
Deception dynamics are introduced by modulating exaggeration using a scalar time schedule $u(t)\!\in[0,1]$ through the weights
\begin{equation}
w_1(t) = \kappa_0\,\big(1-u(t)\big),\qquad
w_2(t) = \alpha_0\,u(t)
\label{eq:weights}
\end{equation}
with constants $\alpha_0\!>\!0$ (deception gain) and $\kappa_0\!>\!0$ (true-goal gain). The single-step instantiation of \eqref{eq:cost_function} is now
\begin{equation}
\mathcal{C}_t \;\equiv\; \mathcal{C}(\Theta_t,\Gamma_t) 
\label{eq:specific_cost_def}
\end{equation}
where
\begin{equation}
\begin{aligned}
\mathcal{C}_t = {} & w_1(t) C_{\mathrm{goal}}(G, s_t, s_{t+1}) \\
& + w_2(t) C_{\mathrm{decep}}(G, F, s_t, s_{t+1}) .
\end{aligned}
\label{eq:specific_cost}
\end{equation}

We adopt a piecewise-linear scheduling profile for the deception weight. Let $S\!\ge\!0$ be the switch-on time, $L\!>\!0$ be the total scheduling window length, and let $r_{\mathrm{up}}\!\ge\!0$ and $r_{\mathrm{down}}\!\ge\!0$ denote, respectively, the ramp-up and ramp-down durations. The schedule is
\begin{equation}
\label{eq:schedule}
u(t)=
\begin{cases}
0, & t < S,\\[2pt]
\dfrac{t-S}{r_{\mathrm{up}}}, & S \le t < S+r_{\mathrm{up}},\\[10pt]
1, & S+r_{\mathrm{up}} \le t < S+L-r_{\mathrm{down}},\\[2pt]
\dfrac{S+L-t}{r_{\mathrm{down}}}, & S+L-r_{\mathrm{down}} \le t < S+L,\\[8pt]
0, & t \ge S+L.
\end{cases}
\end{equation}

\begin{figure}[H]
\includegraphics[width=0.99\textwidth]{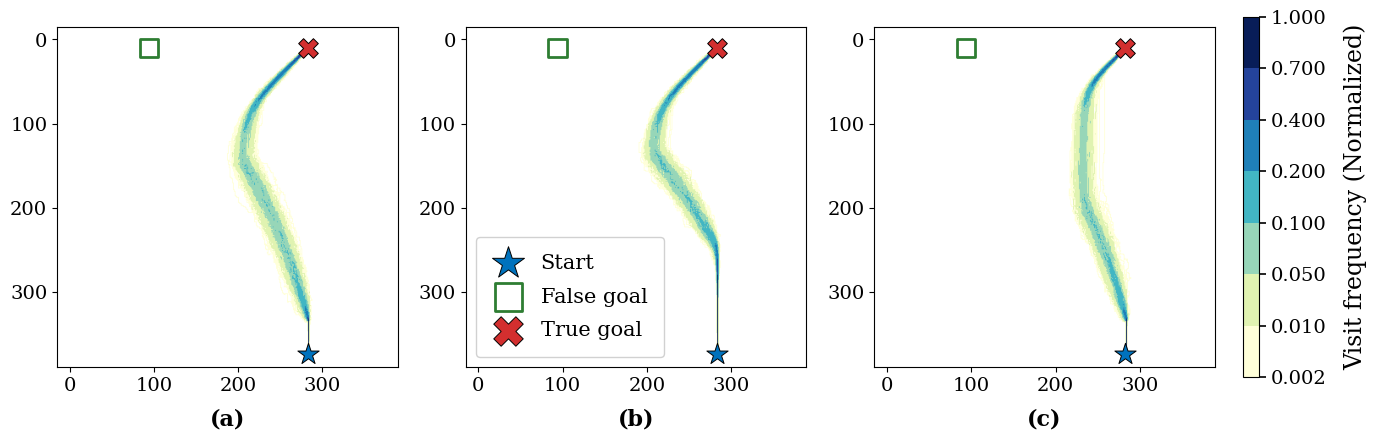}
\caption{Exaggeration: Single-agent trajectory heatmaps over 500 trials. Three exaggeration deception scheduling cases are shown; (\textbf{a}) Hard Switch, (\textbf{b}) Ramp Up, and (\textbf{c}) Ramp Down. These illustrate dynamically tuning the deceptive trajectories from the start (star) towards the false goal $F$ (square) before transitioning to the true goal $G$ (red cross).}
\label{fig:sa-heat}
\end{figure}

Additional parameters were set as follows; $\alpha_0=4$, $\kappa_0=10$, $\lambda=0.8$, and lookahead depth $K_t=3$. 
{
For the single-agent exaggeration experiment, we use a one-step cost formulation combined with a multi-step planning horizon. Along each trajectory, 
the cost is updated at each step based on the state, and it is accumulated over the trajectory.} 
The temporal constants are $S=40$ and $L=300$. 
We consider three cases in (\ref{eq:schedule}) governed by the ramp parameters $(r_{\mathrm{up}}, r_{\mathrm{down}})$,  
set to $(0,0)$ for Hard Switch, $(100,0)$ for Ramp Up, and $(0,100)$ for Ramp Down.
The three cases are illustrated in Figure~\ref{fig:sa-heat}, plotting trajectory heat maps over 500 trials.

As illustrated in Figure~\ref{fig:sa-heat}, Algorithm~\ref{alg:single_agent_dpp} 
effectively generates exaggerated trajectories towards the false goal in the upper left.
The differences between the three cases
arise directly from the temporal weighting of deception defined in \eqref{eq:weights}.  
The Ramp Up case eventually commits to a more exaggerated path, whereas Ramp Down has a strong early bias to exaggeration.
In comparison, the Hard Switch trajectories are focused on the false goal and then abruptly switch to the true~goal. 

\vspace{-3pt}

\subsubsection{Ambiguity}\label{sec:sa-ambiguity}
Next, we consider deception through ambiguity. The agent deviates from the most efficient path to obscure the true goal by using a trajectory that is balanced between the true and false goal(s), e.g., as defined 
in~\cite{savas2021}. The agent progresses more or less equally towards both possible goals, so that the observer cannot readily predict which is the true goal until the agent eventually closes on the true goal. 

All settings (grid, action set, receding-horizon loop, and Boltzmann sampling) are identical to the exaggeration case above; only the cost differs. To maintain ambiguity, we penalize the absolute distance difference between the false and true goals so that the agent tends to approach and stay near a line of bisection. Along the ideal bisector line, the Euclidean distance to either goal remains the same.  
With Euclidean distance $d(\cdot,\cdot)$ and a one-step candidate $s_{t+1}$, we obtain
\begin{equation}
C_{\mathrm{amb}}(G,F,s_{t+1}) \triangleq \big|\, d(s_{t+1},F) - d(s_{t+1},G) \,\big|. 
\label{eq:amb_cost_component}
\end{equation}
Balancing ambiguity with progress toward the true goal (using the same goal-progress term $C_{\mathrm{goal}}(G,s_t,s_{t+1})$ as before), the per-step total cost is now
\begin{equation}
\mathcal{C}_t \;=\; \tilde w_1(t)\, C_{\mathrm{amb}}(G,F,s_{t+1})
\;+\; \tilde w_2(t)\, C_{\mathrm{goal}}(G,s_t,s_{t+1}) .
\label{eq:amb_total_cost}
\end{equation}

We adopt the time scheduling approach as before, where now the gains for the ambiguity objective are set to $\alpha=4$ and $\kappa=4$, and the temporal constants are adjusted to $S=40$ and $L=250$ to accommodate the specific dynamics of intent concealment. 
The ramp parameters $r_{\mathrm{up}}$ and $r_{\mathrm{down}}$ are the same as those used above. 

\begin{figure}[H]
    \includegraphics[width=0.98\textwidth]{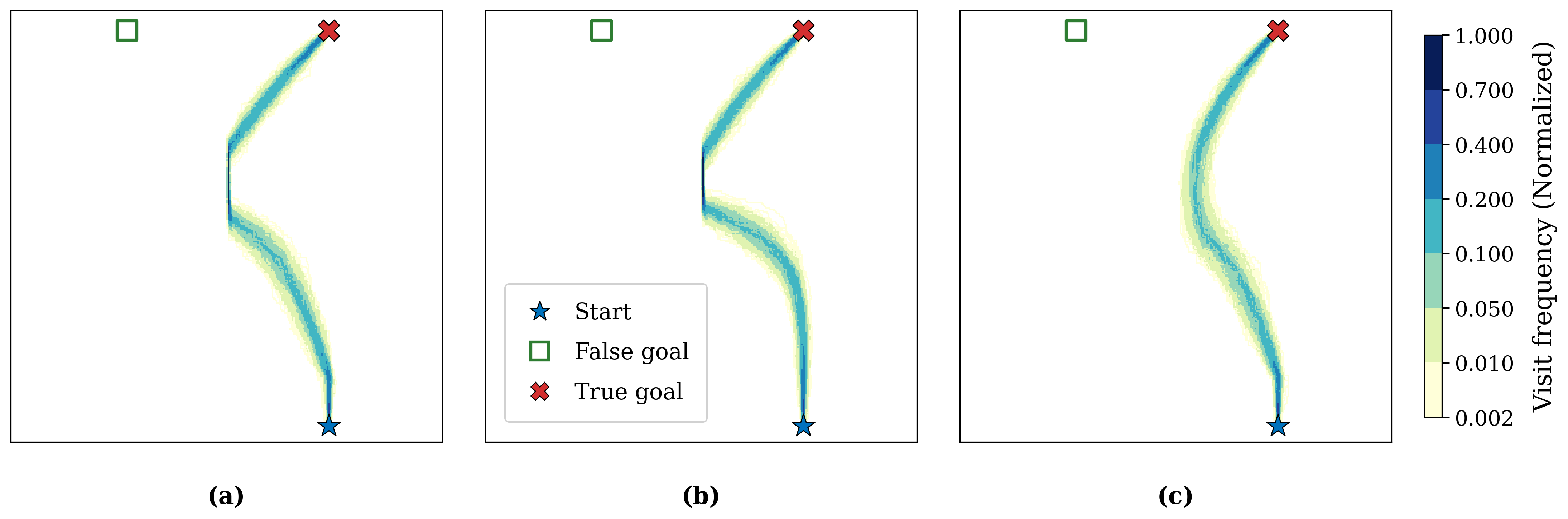}
\caption{Ambiguity: Single-agent trajectory heatmaps over 500 trials. Three ambiguity deception scheduling cases are shown; (\textbf{a}) Hard Switch, (\textbf{b}) Ramp Up, and (\textbf{c}) Ramp Down. These illustrate dynamically tuning the ambiguity deception from the start (star) by approaching the bisection line where the true and false goals are equi-distant, before transitioning to the true goal $G$ (red cross). }
    \label{fig:sa-amb}
\end{figure}

Ambiguity trajectory heatmaps over 500 trials are shown in 
Figure~\ref{fig:sa-amb}.  Similar to Figure~\ref{fig:sa-heat}, we let $\tilde w_1(t)$ and $\tilde w_2(t)$ follow the same three-piecewise schedulers used previously.
Now, the temporal schedule balances the ambiguity cost $C_{\mathrm{amb}}$ against direct progress to the true goal. 
Depending on the dynamic weighting schedule, the trajectories progress towards the line of bisection and maintain ambiguity, then transition towards the true goal. 
All three cases show the agent progressing to the line of bisection as the agent moves from bottom to top. Along this line, the agent's distance to both the true and false goals remains nearly constant.  Depending on the schedule, the agent eventually turns toward the true~goal.

\subsubsection{Obstacles and Trajectory Alignment}
\label{obstacles}

\textls[-15]{Our proposed Boltzmann-based Algorithm~\ref{alg:single_agent_dpp} is designed to maintain a consistent global intent while providing the necessary flexibility for local adaptation during execution. This allows the agent to, for example, navigate around unexpected obstacles without compromising its deceptive objective. The proposed approach can be integrated into standard navigation pipelines, 
heatmap-guided exploration, or safety-zone--constrained mission planning. Local environmental 
updates (e.g., temporary obstacles, risk zones, or dynamic no-go regions) can be 
incorporated without resorting to a new global planning process.}

To implement the local constraints while retaining the full state space, we impose an infinite cost for transitions into obstacle locations, yielding
\begin{equation}
\begin{aligned}
\mathcal{C}_t = {} & w_1(t) C_{\mathrm{goal}}(G, s_t, s_{t+1}) \\
& + w_2(t) C_{\mathrm{decep}}(G, F, s_t, s_{t+1}) \\
& + \infty \cdot \mathbb{I}(s_{t+1}),
\end{aligned} 
\label{eq:specific_cost_obstacles}
\end{equation}
where $\mathbb{I}(\cdot)$ is the indicator function given by
\begin{equation}
\mathbb{I}(s_{t+1}) = 
\begin{cases} 
1 & \text{if transition into } s_{t+1} \text{ is infeasible} \\ 
0 & \text{otherwise} .
\end{cases}
\end{equation} 

\begin{figure}[H]
    \includegraphics[width=0.6\columnwidth]{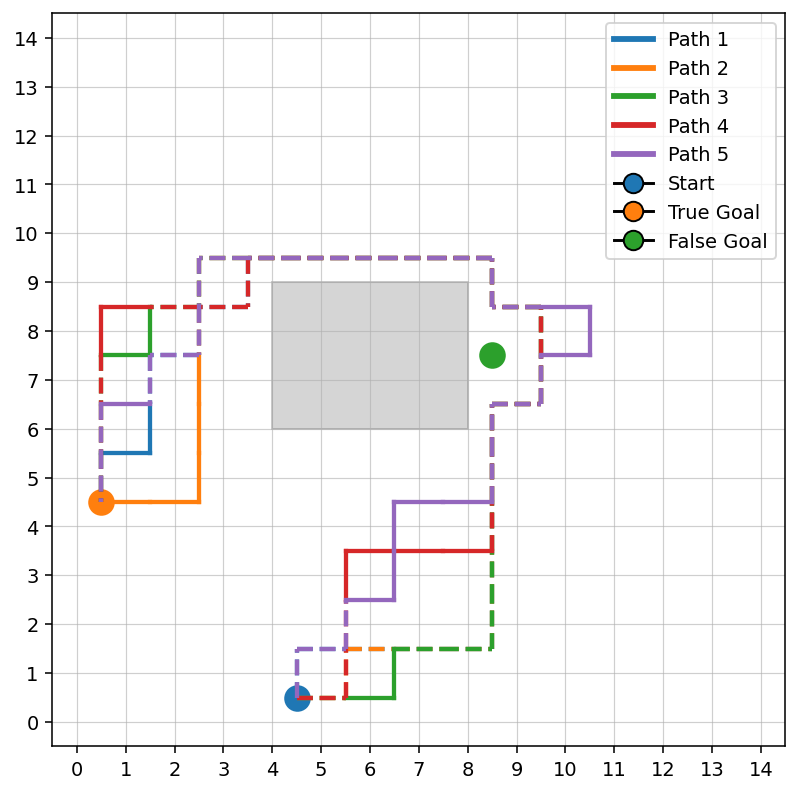}
    \caption{Examples of exaggeration paths. The stochastic policy maintains deception while avoiding the local obstacle region (in gray).}
    \label{fig:F_obstacle}
\end{figure}

Sample trajectories in Figure~\ref{fig:F_obstacle} demonstrate exaggeration while bypassing an obstacle. The same stochastic policy generates all paths shown.

To further characterize deception along a trajectory, such as those in Figure~\ref{fig:F_obstacle}, we quantify deceptive steering using a continuous angular metric 
inspired by the trajectory--alignment perspective introduced in~\cite{gutierrez2025dppp}. 
Let $\theta^{(a)}_{\upsilon}$ and $\theta^{(d)}_{\upsilon}$ denote the angular deviations between the agent’s instantaneous motion direction and the directions to the true and decoy goals at time step $\upsilon$.
We define the continuous alignment score (CAL) as
\begin{equation}
\label{eq:CAL}
\mathrm{CAL} = \frac{1}{N_{\mathrm{CAL}}} \sum_{\upsilon=1}^{N_{\mathrm{CAL}}} \big( \theta^{(a)}_{\upsilon} - \theta^{(d)}_{\upsilon} \big) .
\end{equation}
\textls[-20]{The angular deviations $\theta_v^{(a)}$ and $\theta_v^{(d)}$ are measured in radians. Under the sign convention in~\eqref{eq:CAL}, negative CAL values indicate that the agent is more closely aligned with the decoy direction, whereas positive values indicate closer alignment with the true-goal direction. 
The sign and magnitude of CAL provide a directed measure of deceptive intent. Negative values indicate orientation toward the decoy, and the absolute value reflects the strength of this angular bias.  Similarly, positive values indicate more emphasis towards the true goal direction. 
A persistently negative CAL suggests a deliberate and consistent behavioral bias that prioritizes misleading the observer over path efficiency.
This metric is relatively insensitive to small random zig-zag angular variations over the trajectory, as they tend to average out.} 

\begin{figure}[H]
    \includegraphics[width=0.65\linewidth]{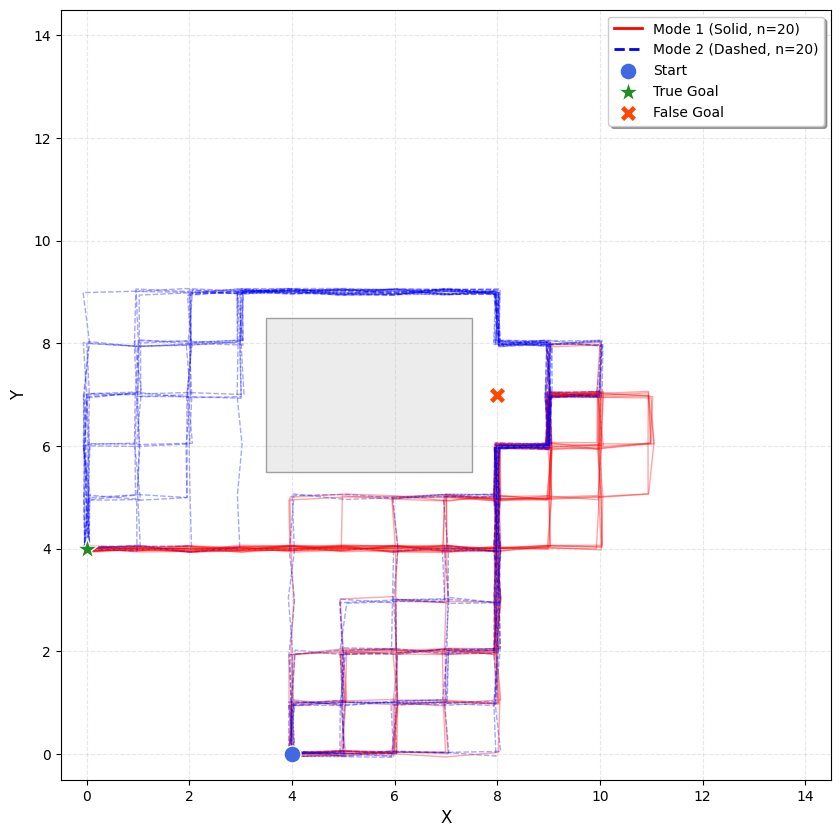}
    \caption{
Forty exaggeration sample trajectories, with blue and red corresponding to 20 randomly selected within each CAL mode shown in Figure \ref{fig:cal_distribution}. The same random policy generates trajectories that are persistently deceptive while avoiding the obstacle. }
    \label{fig:obstacle_paths}
\end{figure}

\begin{figure}[H]
    \includegraphics[width=0.85\linewidth]{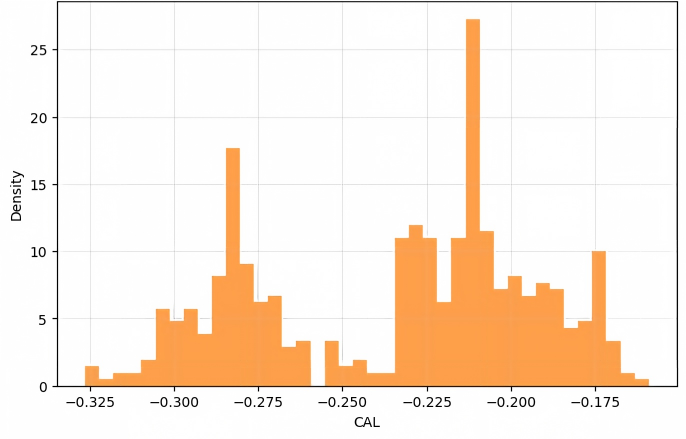}
    \caption{
Example empirical distribution of the {Continuous Alignment score (CAL)} obtained
from 500 exaggerated deception trajectory trials. The two CAL modes are illustrated in Figure~\ref{fig:obstacle_paths}. 
    }
    \label{fig:cal_distribution}
\end{figure}

The use of CAL is illustrated in  Figures~\ref{fig:obstacle_paths} and \ref{fig:cal_distribution}.  
An example empirical distribution of CAL is shown in Figure~\ref{fig:cal_distribution}, obtained from 500 trials of exaggeration deception; note the two clusters (modes). All trajectories yield negative CAL, as expected with exaggeration. 
These trajectories are drawn from the scenario shown in Figure~\ref{fig:obstacle_paths}.
In this figure, 40 exaggeration trajectories are plotted, with blue and red colors each corresponding to 20 randomly chosen trajectories from within each CAL mode. 
Larger negative CAL values indicate a trajectory that more aggressively exaggerates towards the false goal for a longer duration while also avoiding the gray obstacle area and maintaining the deceptive trajectory.

\subsubsection{Comparison with End-to-End DPP}

\begin{figure}[H]

    \begin{minipage}[b]{0.48\columnwidth}
        \centering
        \includegraphics[width=\linewidth]{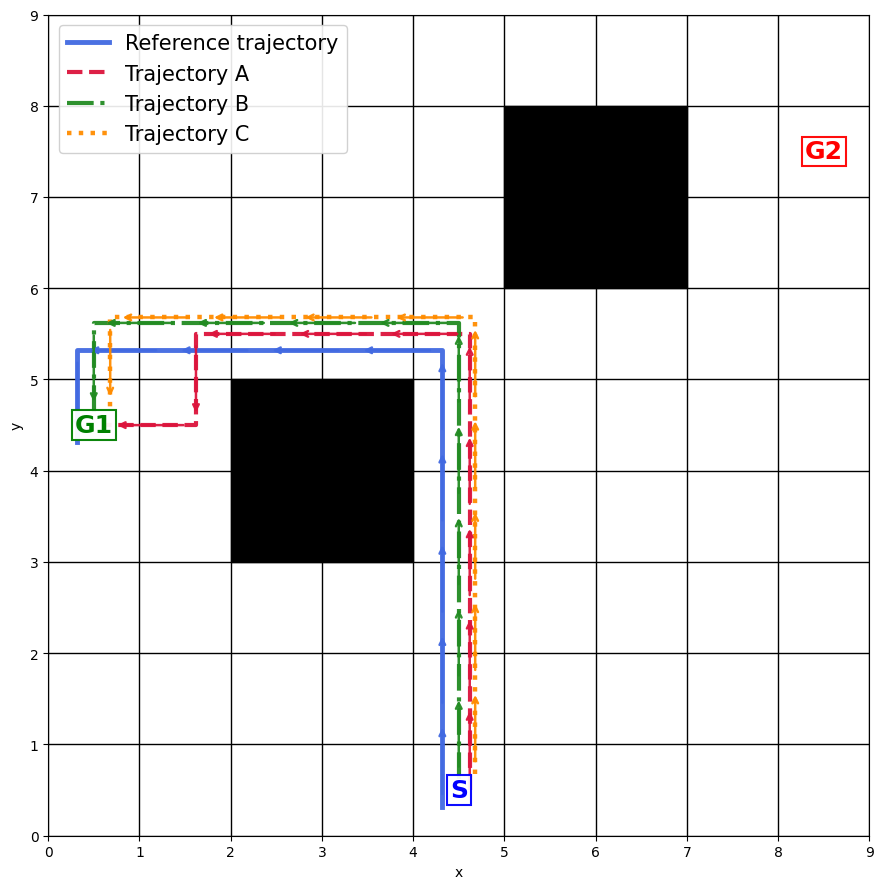}
        \caption*{(\textbf{a})}
    \end{minipage}
    \hspace{6pt}
    \begin{minipage}[b]{0.48\columnwidth}
        \centering
        \includegraphics[width=\linewidth]{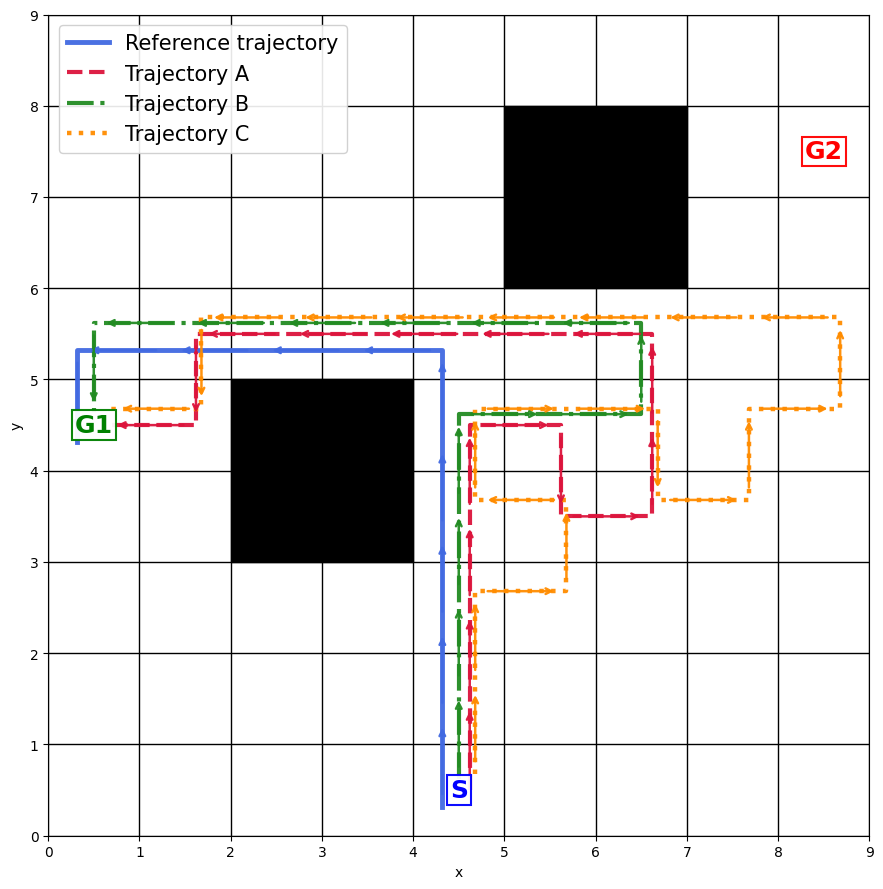}
        \caption*{(\textbf{b})}
    \end{minipage}

\caption{Exaggeration deception example, comparing an end-to-end plan using the LP optimization in~\cite{savas2021} with random trajectories generated by the proposed approach. The setup corresponds to Figure~2 in~\cite{savas2021}, with true goal G1 and false goal G2. The LP method yields a single reference trajectory, shown in both panels.  (\textbf{a}) $\lambda=12$. With low randomness, realizations of our method track the LP-based plan.  
(\textbf{b}) $\lambda=4$. Increased randomness yields trajectories that have more deviation from the LP-based reference trajectory. 
}
    \label{fig:two-panel-single-column}
\end{figure}

Next, we compare our planner with an end-to-end reference trajectory generated by the LP-based method from~\cite{savas2021} using exaggeration deception. 
Results are shown in Figure~\ref{fig:two-panel-single-column}, where the grid setup corresponds to the example shown in Figure 2 of~\cite{savas2021}, with the true goal marked G1 and the false goal marked G2. 
We tune our method using parameters $\alpha_0=2$, $\kappa_0=8$, $K_t=3$, $S=0$, and $L=4$.
We show two cases with different choices for the parameter $\lambda$ that induces randomness in our Boltzmann-based policy, compared to the same reference trajectory.
With low randomness, shown in  Figure~\ref{fig:two-panel-single-column}a, our method tracks the end-to-end solution with little variation among realizations.  With higher randomness, shown in  Figure~\ref{fig:two-panel-single-column}b, our method exhibits more variation with respect to the \linebreak  reference trajectory.

Although our method does not explicitly optimize the LP objective in an end-to-end fashion, in this simple example, our method is shown to replicate the LP-based trajectory.
While all trajectories exhibit the desired exaggeration behavior, initially exaggerating toward the false goal before converging to the true goal, tuning $\lambda$ enables a narrower or wider spread of realizations.

\subsection{Dynamic Deception: Moving False Goal}

Our method enables new forms of dynamic deception by time-varying the agents' parameters. As an illustrative example, we let the true goal be static and move the false goal along a linear trajectory, while executing a deceptive trajectory.

\begin{figure}[H]

    \subfloat[Ambiguity, slow false goal.\label{subfig:amb_slow}]{
        \includegraphics[width=0.46\textwidth]{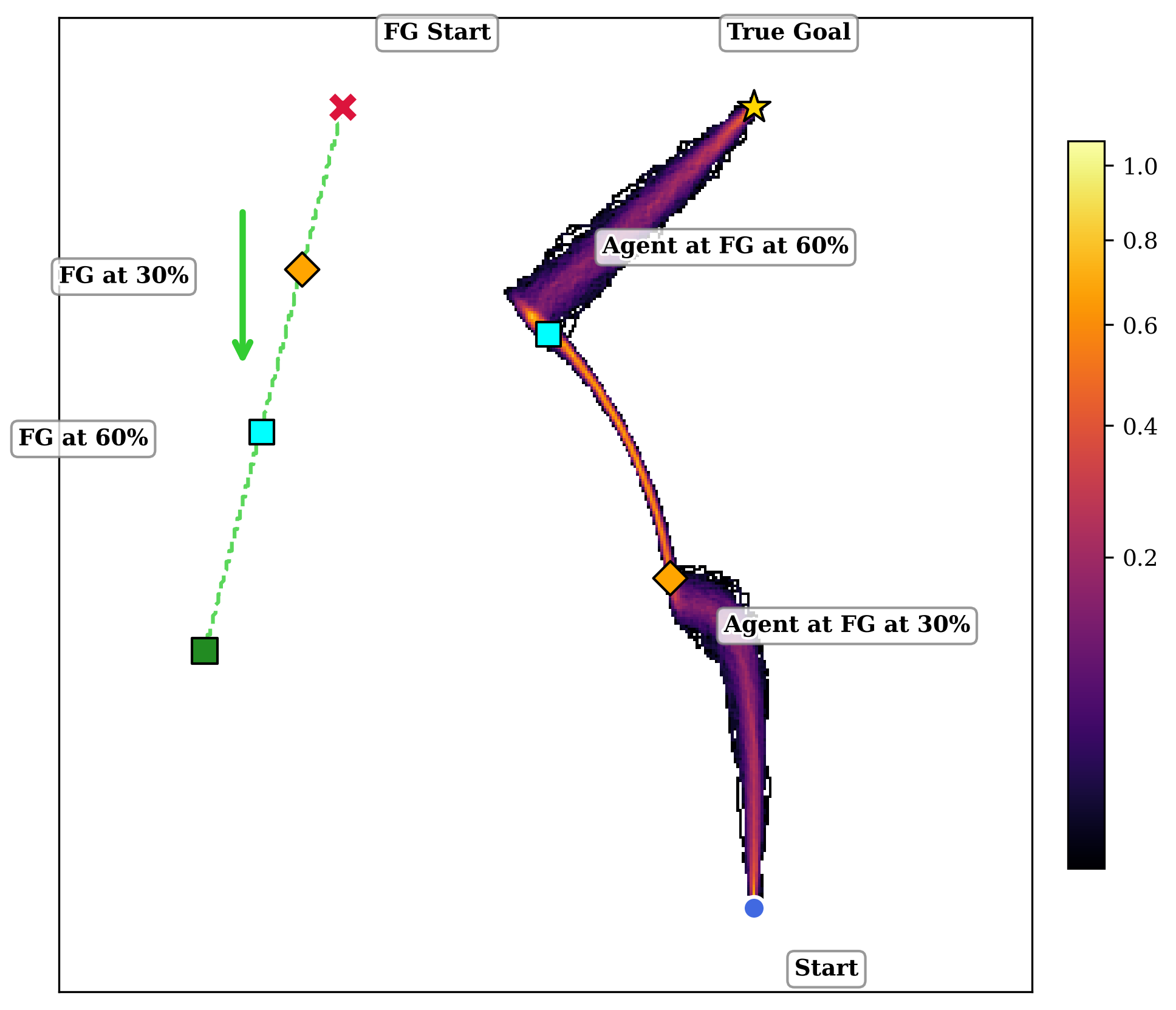}
    }\hfill
    \subfloat[Ambiguity, fast false goal.\label{subfig:amb_fast}]{
        \includegraphics[width=0.46\textwidth]{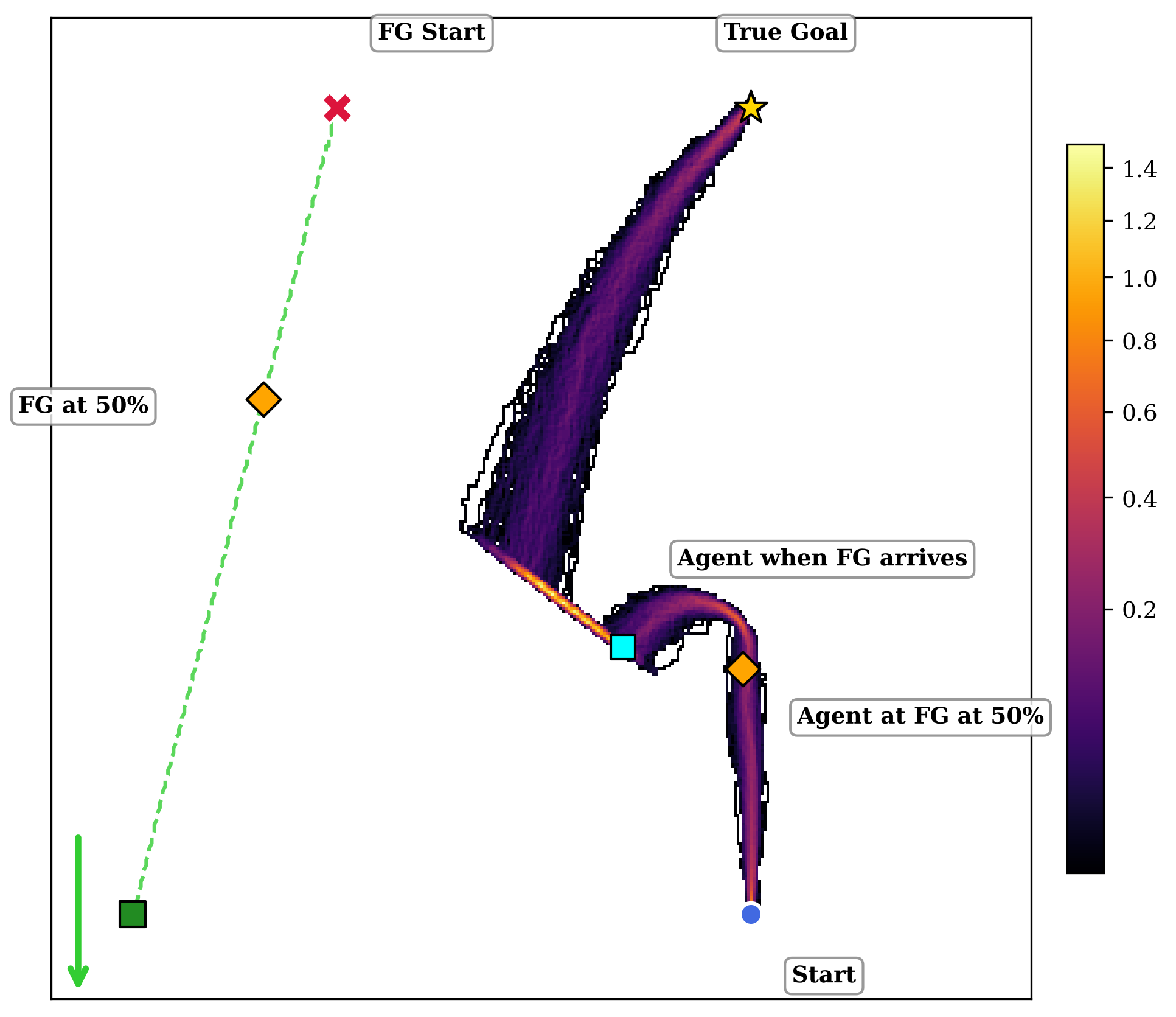}
    }

    \vspace{0.5em}

    \subfloat[Exaggeration, slow false goal.\label{subfig:exa_slow}]{
        \includegraphics[width=0.46\textwidth]{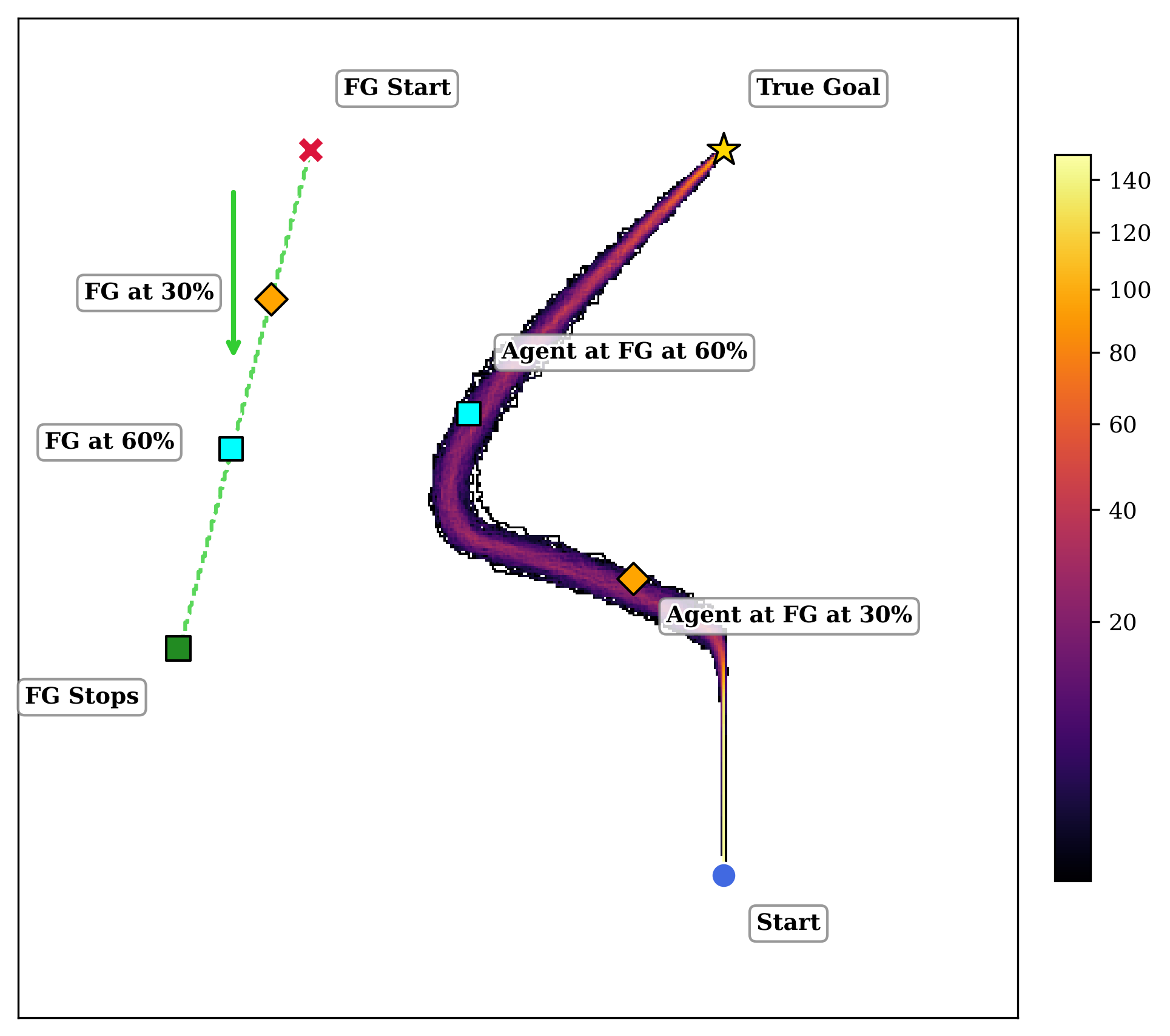}
    }\hfill
    \subfloat[Exaggeration, fast false goal.\label{subfig:exa_fast}]{
        \includegraphics[width=0.46\textwidth]{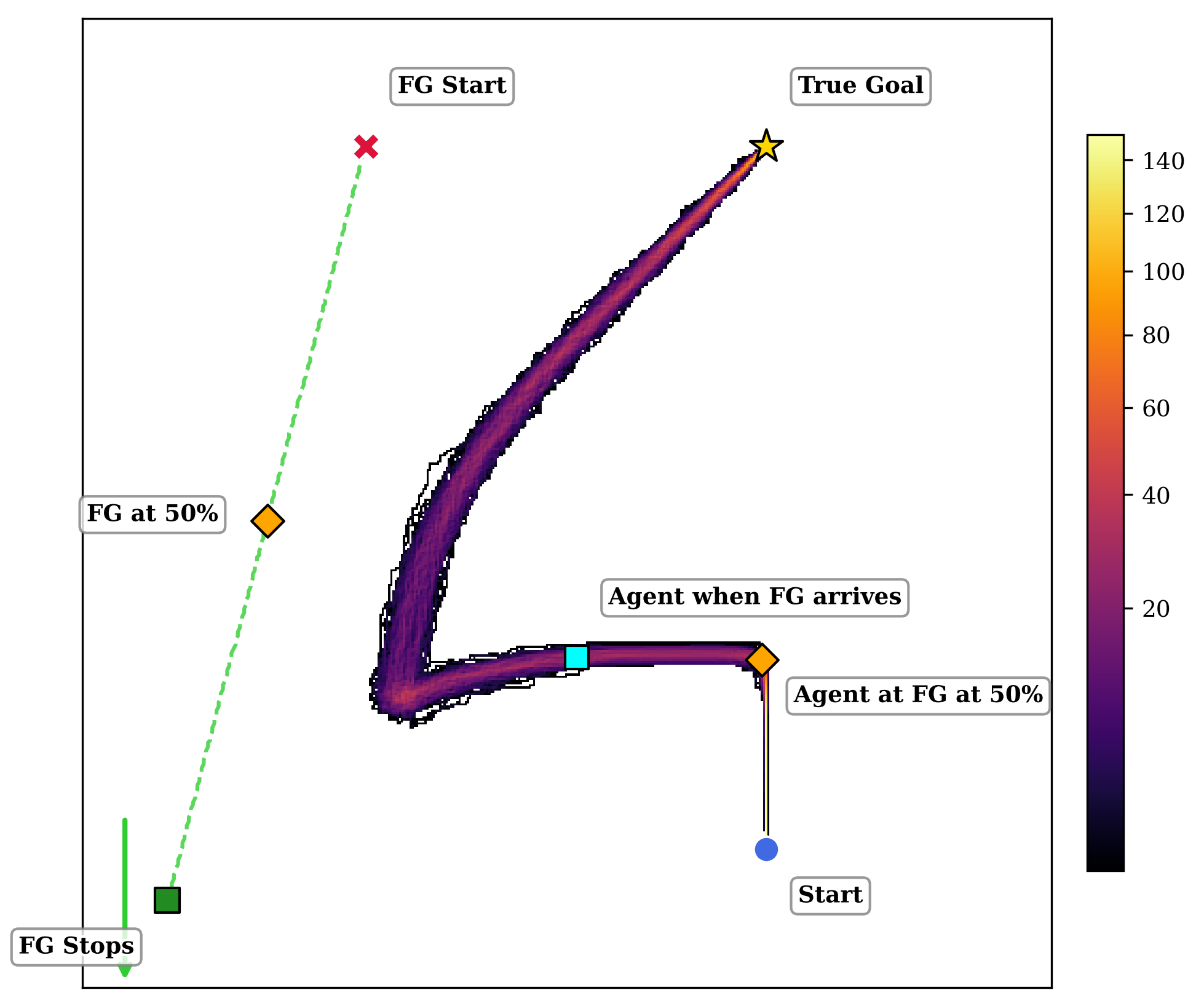}
    }

    \caption{Deception with a moving false goal. Each panel shows 500 trials. Ambiguity, panels \mbox{(\textbf{a},\textbf{b}), and} exaggeration, panels (\textbf{c},\textbf{d}), with slow and fast false goals moving from top-to-bottom. Deceptive path planning dynamically responds to the moving false goal before eventually turning toward the true goal.}
    \label{fig:combined_deception}
\end{figure}

Consider the scenario shown in Figure~\ref{fig:combined_deception}, where the agent starts at the bottom right with the true goal in the upper right, and the false goal moves linearly from upper left to bottom left.
Two moving false goal cases are shown: slow (0.4) and fast (2.0) grid units per time step. Both the agent and the false goal begin motion simultaneously. 
We apply ambiguous deception in Figure~\ref{fig:combined_deception}a,b, 
and exaggeration deception in Figure~\ref{fig:combined_deception}c,d, plotting a heat map of 500 trials for each case.
For ambiguity, we use the same agent parameters, including $u(t),$ as in Figure~\ref{fig:sa-amb}, and 
for exaggeration, we use the same parameters as in Figure~\ref{fig:sa-amb}.

\vspace{-9pt}

With ambiguity deception shown in Figure~\ref{fig:combined_deception}a,b, as the false goal moves, so does the bisector between the false and true goal. The agent maintains ambiguity by tracking the moving bisector before eventually committing to the true goal.
With exaggeration deception in 
Figure~\ref{fig:combined_deception}c,d, the agent initially steers toward the moving false goal, with the direction of deceptive movement tracking the false goal movement. Ultimately, the agent redirects toward the true goal.

These examples illustrate the dynamic flexibility of our approach, and many interesting behaviors may be induced through various parameter tunings, including adding and dropping goals and false goals, and altering the form of deception and its emphasis. 
Note also that our method can be generalized to target pursuit with deception, where the moving goal is the perceived target, and a virtual moving false goal is employed to confuse the target by not following an optimal pursuit path. 
\subsection{Multi-Agent Deception with Joint Budget Constraint} \label{sec:problem}

Next, we present a multi-agent example, where a joint constraint induces coupling among the agents. 
We consider a multi-agent team of $N$ cooperative agents $\mathcal{I}=\{1,\dots,N\}$ moving on a graph $(\mathcal{V},\mathcal{E})$ with neighborhood map $\mathcal{N}:\mathcal{V}\!\to\!2^{\mathcal{V}}$. The team seeks to reach a true goal set $\mathcal{G}_T \subseteq \mathcal{V}$ while feigning a false goal set $\mathcal{G}_F \subseteq \mathcal{V}$ to deceive an observer. Agent $i \in \mathcal{I}$ follows a path $\tau_i = (s^i(0), \dots, s^i(T))$, where $s^i(t+1) \in \mathcal{N}(s^i(t))$. Let $\boldsymbol{\tau} \triangleq (\tau_1, \dots, \tau_N)$ denote the joint path, and 
{$\boldsymbol{\phi}(\cdot)$ denotes the coupled cost term. Following \eqref{eq:individual_cost_function} and \eqref{eq:interaction_cost_vector},
the time-indexed cost at each step $t$ is
\begin{equation}
\label{eq:Ct}
\begin{aligned}
\mathcal{C}_t \;\equiv\; \mathcal{C}\!\big(\Theta_t,\Gamma_t\big)
&= -\sum_{i \in \mathcal{I}} \kappa_i\,\Delta^{\text{prog}}_i(t) \\
&\quad - \sum_{i \in \mathcal{I}} \beta_i\,\Delta^{\text{dec}}_i(t)
      + \boldsymbol{\phi}\big(S(t)\big) ,
\end{aligned}
\end{equation}
where for each agent $i \in \mathcal{I}$ we denote $\kappa_i$ and $\beta_i$ as the weight coefficients for progress to the true goal and exaggeration to a false goal, respectively. 
}

A standard measure of progress to the true goal $G^i(t)$ is the one-step distance drop,
\begin{equation}
\label{eq:progress}
\Delta^{\text{prog}}_i(t) = \max \left\{ 0, d(s^i(t), G^i(t)) - d(s^i(t+1), G^i(t)) \right\}
\end{equation}
and considering we apply exaggeration as the deception form, a convenient exaggeration term toward the false goal $F^i(t)$ is then
\begin{equation}
\label{eq:deception}
\Delta^{\text{dec}}_i(t) = \max \left\{ 0, d(s^i(t), F^i(t)) - d(s^i(t+1), F^i(t)) \right\}.
\end{equation}
Note that both \eqref{eq:progress} and \eqref{eq:deception} are non-negative. More progress per step is desired, so negative signs appear in  \eqref{eq:Ct} to incentivize larger steps by reducing these costs.

Let the team of agents share a total joint step budget $B_{\text{team}} > 0$, representing the total resources (a proxy for energy) available for motion over the fixed time horizon $T$. For each agent $i \in \mathcal{I}$, the per-step resource consumption is denoted by $b_i\big(s^i(t), s^i(t+1)\big) \ge 0$. The cumulative consumption across all agents is thus restricted by the shared team capacity
\begin{equation}
\label{eq:budget-consumption}
\sum_{t=0}^{T-1} \sum_{i \in \mathcal{I}} b_i\big(s^i(t), s^i(t+1)\big) \le B_{\text{team}} .
\end{equation}

Now, the multi-agent DPP is posed as minimizing the sum of local per-step costs while satisfying the total budget, 
\begin{equation}
\label{eq:team-program}
\begin{aligned}
\min_{\boldsymbol{\tau}} \quad & \sum_{t=0}^{T-1} \mathcal{C}_t \\
\text{s.t.} \quad & s^i(t+1) \in \mathcal{N}\big(s^i(t)\big), \quad \forall i \in \mathcal{I}, \forall t \\
& \sum_{t=0}^{T-1} \sum_{i \in \mathcal{I}} b_i\big(s^i(t), s^i(t+1)\big) \le B_{\text{team}} .
\end{aligned}
\end{equation}

This explicitly incorporates the joint budget-limiting inequality constraint to enforce the team's resource capacity. 

At each time step $t$, the next joint move $S(t+1)$ is selected from the set of feasible joint actions $\mathcal{A}(t)$ using the Boltzmann policy 
with the per-step cost $\mathcal{C}_t$. Feasible joint paths are restricted to those that maintain the shared budget constraint~\eqref{eq:budget-consumption} while ensuring sufficient remaining capacity for agents to reach the true goals.

\begin{figure}[H]
  \setlength{\abovecaptionskip}{3pt}
  \setlength{\belowcaptionskip}{0pt}

  \begin{minipage}[b]{0.47\linewidth}
    \centering
    \includegraphics[width=\linewidth]{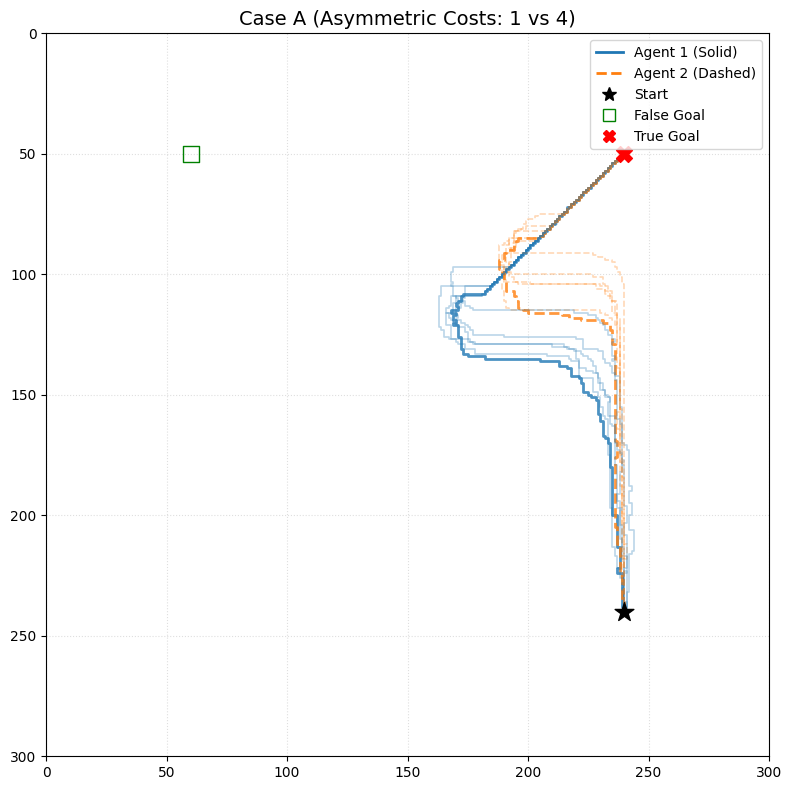}
    \caption*{(\textbf{a}) Different budgets $(1,4)$}
  \end{minipage}\hfill
  \begin{minipage}[b]{0.47\linewidth}
    \centering
    \includegraphics[width=\linewidth]{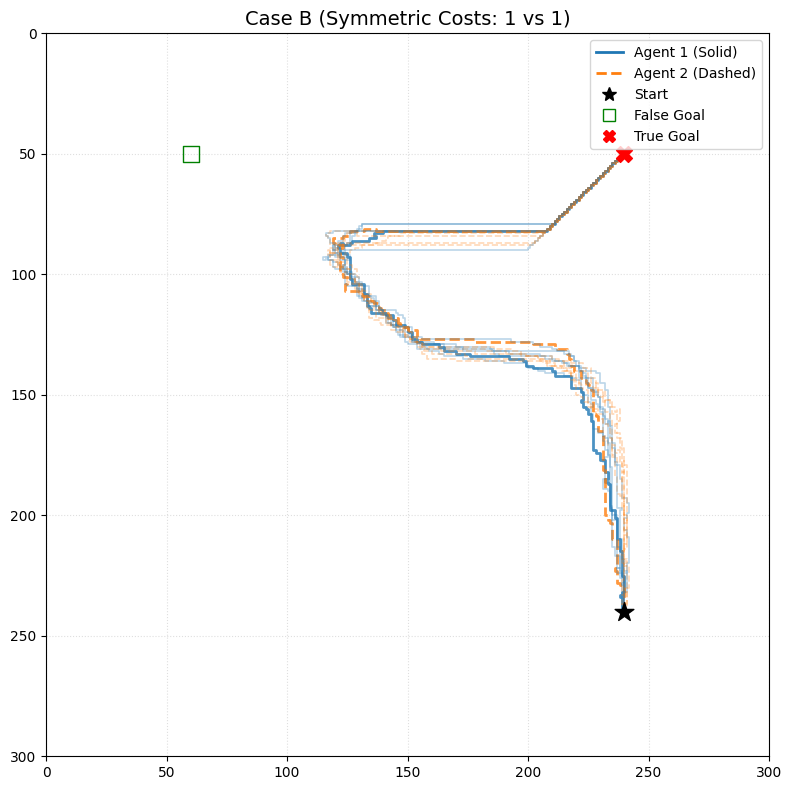}
    \caption*{(\textbf{b}) Identical budgets $(1,1)$}
  \end{minipage}

  \vspace{0.1em}

  \begin{minipage}[b]{0.47\linewidth}
    \centering
    \includegraphics[width=\linewidth]{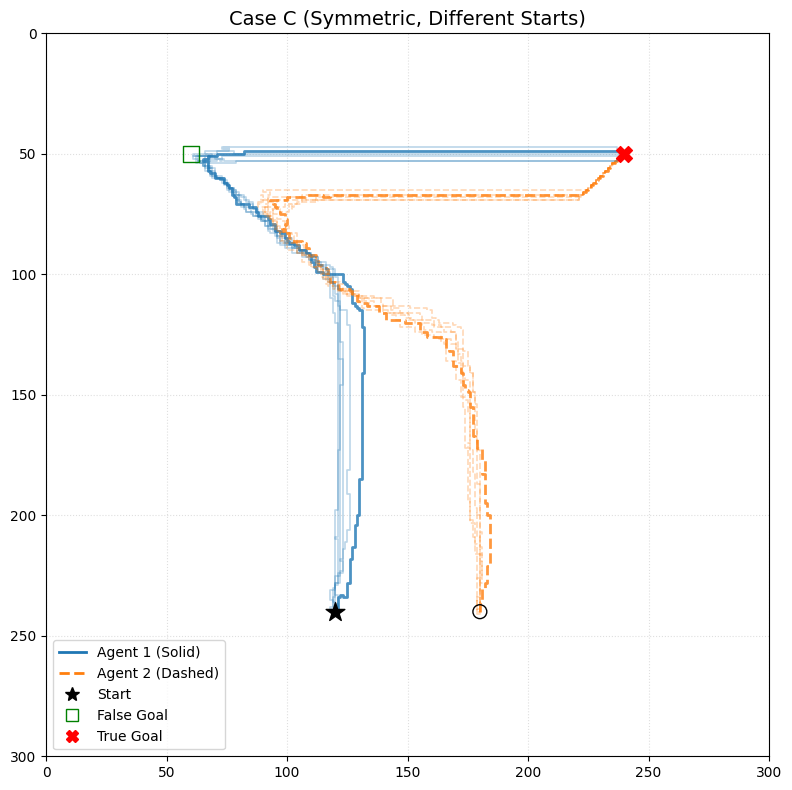}
    \caption*{(\textbf{c}) Identical budgets $(1,1)$}
  \end{minipage}\hfill
  \begin{minipage}[b]{0.47\linewidth}
    \centering
    \includegraphics[width=\linewidth]{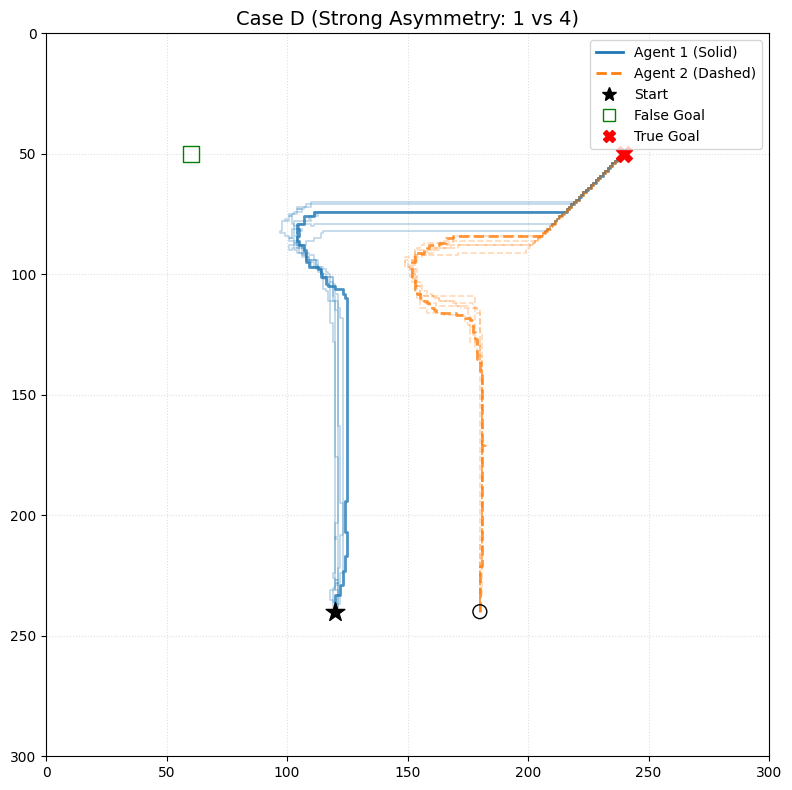}
    \caption*{(\textbf{d}) Different budgets $(1,4)$}
  \end{minipage}

  \vspace{0.1em}

  \caption{
  Two-agent deception example with total move joint budget constraints. (\textbf{a}) Different relative agent budgets $(1,4)$, same start location. (\textbf{b}) Same agent budgets $(1,1)$, same start. (\textbf{c}) Same agent budgets, different starts. (\textbf{d}) Different budgets $(1,4)$, different starts. Higher agent budgets enable more deceptive path variation with respect to the most efficient start-to-goal path.
  }
  \label{fig:random-realizations-row}
\end{figure}

We instantiate a two-agent ($N=2$) joint Boltzmann planner under the shared-capacity budget formulation (Figure \ref{fig:random-realizations-row}). At each time $t$, a joint action is sampled with $\lambda=0.6$ from the feasible set $\mathcal{A}(t)$, which is pruned to ensure the cumulative non-optimal expenditure $D_{\text{used}}$ never exceeds the deception capacity $D_{\max}$. For this setup, we set $B_{\text{team}} = 1500$ and $\gamma = 0.3$, yielding $D_{\max} = 450$ as the ``strategic slack'' extracted from the initial energy reserve. The cost $\mathcal{C}_t$ in \eqref{eq:Ct} is used with $\kappa_i = 1.0$ and $\beta_i(f) = w_{\mathrm{F}}(f)$, where the intensity $w_{\mathrm{F}}(f)$ follows a triangular ramp,
\begin{equation}\label{eq:wf_tri}
w_{\mathrm{F}}(f) = w_{\text{min}} + (w_{\text{max}} - w_{\text{min}}) \max(0, 1 - |2f - 1|) ,
\end{equation}
with $w_{\text{min}}=0.5$ and $w_{\text{max}}=15.0$. The fraction $f = D_{\text{used}}/D_{\max}$ measures the depletion of this finite resource. 
{To regulate efficiency, the coupling term $\boldsymbol{\phi}(S(t))$ is instantiated as a cost on the collective energy increment
\begin{equation}
\label{eq:phi_instantiation}
\boldsymbol{\phi}(S(t)) = \alpha \sum_{i \in \mathcal{I}} \Delta \varepsilon_i(t) ,
\end{equation}
where $\alpha=0.45$ and $\Delta \varepsilon_i(t) = \max\{0, e_i + \Delta d^i_G(t)\}$ represents the net energy cost relative to the goal. 
For this coupled energy cost, we favor joint paths that use less total incremental energy, and consequently, $\boldsymbol{\phi}(S(t))$ is added in \eqref{eq:Ct} without a sign change.
}

Four deceptive path cases are shown in Figure \ref{fig:random-realizations-row} with varying start positions and relative energy budget per-agent. The figure shows (a) asymmetric costs $(1,4)$ with identical starts, (b) symmetric costs $(1,1)$ with identical starts, (c) symmetric costs $(1,1)$ from distinct starts, and (d) asymmetric costs {$(1,4)$} from the geometry of (c) to provide a consistent baseline for comparison. For each case, we plot several random realizations for each agent. In this setup, agents with identical marginal costs share the deceptive burden equally, resulting in similar trajectories. Conversely, an agent with a lower marginal cost possesses a higher effective budget that enables increased deception, leading to a more pronounced deceptive deviation from the efficient path.

\subsection{Coupled Multi-Agent Transportation Example}
\begin{figure}[H]
    \includegraphics[width=0.95\columnwidth]{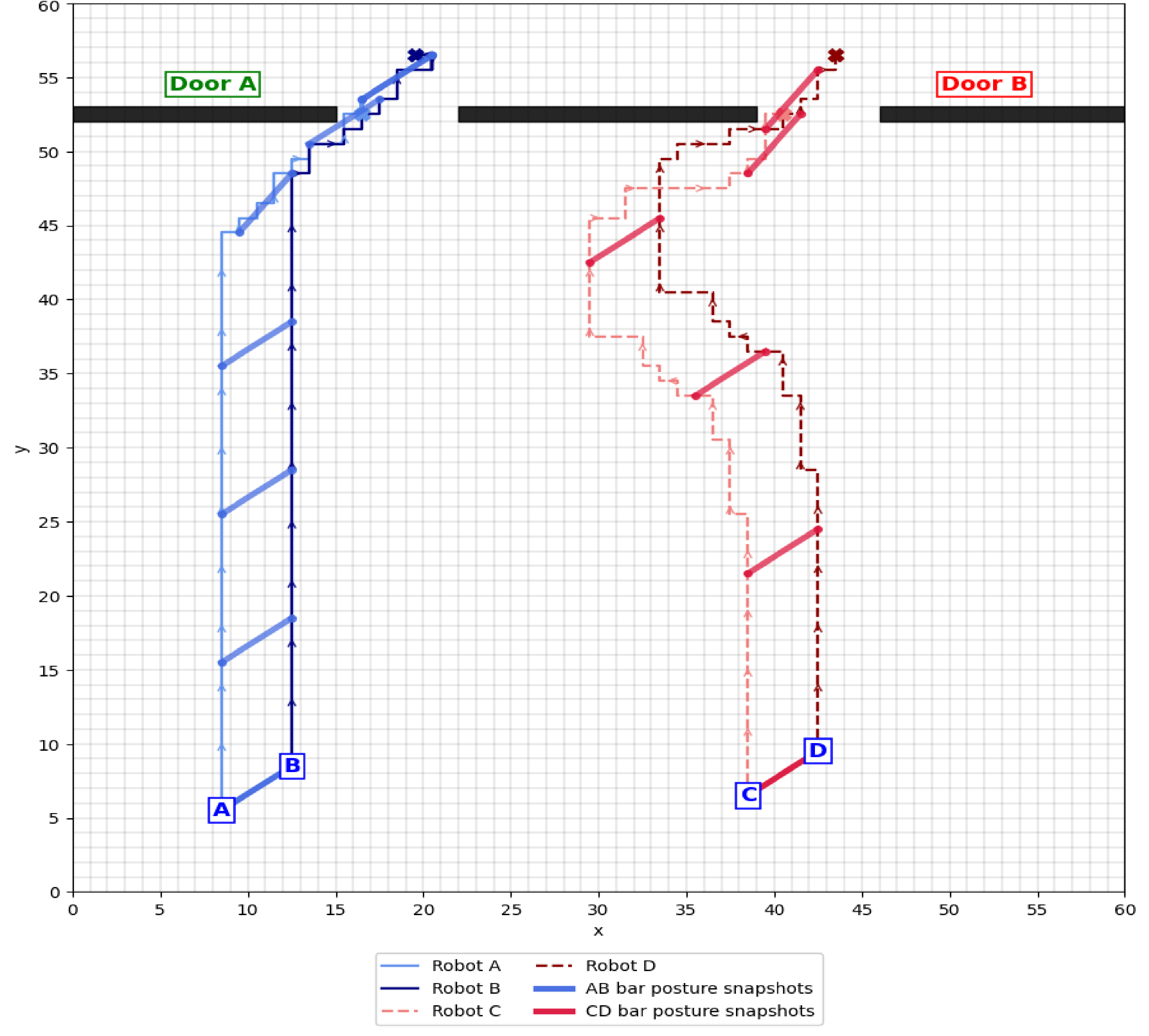}
\caption{Cooperative transportation example with coupled robot motion. Four robots are organized into two teams, each transporting a rigid bar through an assigned narrow door. Team AB basically follows the shortest path to Door A, whereas Team CD first exaggerates toward Door A before redirecting to Door B under a joint Boltzmann policy. The parameters are $\lambda=1.8$, $K_t=3$, $L_b=5$, $\alpha_{CD}=5.5$, $\kappa_{AB}=7.0$, $\kappa_{CD}=7.0$, $S=0$, $L=55$, and $r_{\mathrm{down}}=16$.}
    \label{fig:rigid_bar_example}
\end{figure}

{
We next consider a cooperative transportation scenario with two pairs of robots carrying rigid bars through assigned narrow doors. The setup is shown in Figure~\ref{fig:rigid_bar_example}, with pairs A-B and C-D each transporting a bar.  
Robots A and B transport one rigid bar directly toward "Door A", whereas robots C and D first exaggerate toward Door A before reaching their true goal, "Door B". The trajectories are generated using Algorithm~\ref{alg:multi_agent_dpp}, where a random policy is constructed over the four-agent joint action space. The cost is defined as
\begin{equation}
\label{eq:bar_cost}
\mathcal{C}_t
=
-\kappa_{AB}\Delta^{A}_{AB}(t)
-u(t)\alpha_{CD}\Delta^{A}_{CD}(t)
-\bigl(1-u(t)\bigr)\kappa_{CD}\Delta^{B}_{CD}(t),
\end{equation}
where
\[
\Delta^{A}_{AB}(t)
=
\bigl[d(m_{AB}(t),D_A)-d(m_{AB}(t+1),D_A)\bigr]_+,
\]
\[
\Delta^{A}_{CD}(t)
=
\bigl[d(m_{CD}(t),D_A)-d(m_{CD}(t+1),D_A)\bigr]_+,
\]
and
\[
\Delta^{B}_{CD}(t)
=
\bigl[d(m_{CD}(t),D_B)-d(m_{CD}(t+1),D_B)\bigr]_+,
\]
with $m_{AB}(t)=\bigl(s_A(t)+s_B(t)\bigr)/2$ and $m_{CD}(t)=\bigl(s_C(t)+s_D(t)\bigr)/2$. \textls[-15]{The rigid-bar constraint is enforced by requiring} $\|s_A(t+1)-s_B(t+1)\|=L_b$ and $\|s_C(t+1)-s_D(t+1)\|=L_b$ for every feasible joint action. 
In Figure~\ref{fig:rigid_bar_example}, robots A-B take a shortest (non-deceptive) path to Door A, whereas robots C-D show exaggeration towards the false goal (Door A), before passing through Door B. 
}

\subsection{Volleyball Game: Deceptive Team Attack Path Planning}\label{sec:volleyball}

\begin{figure}[H]
  \includegraphics[width=0.98\columnwidth]{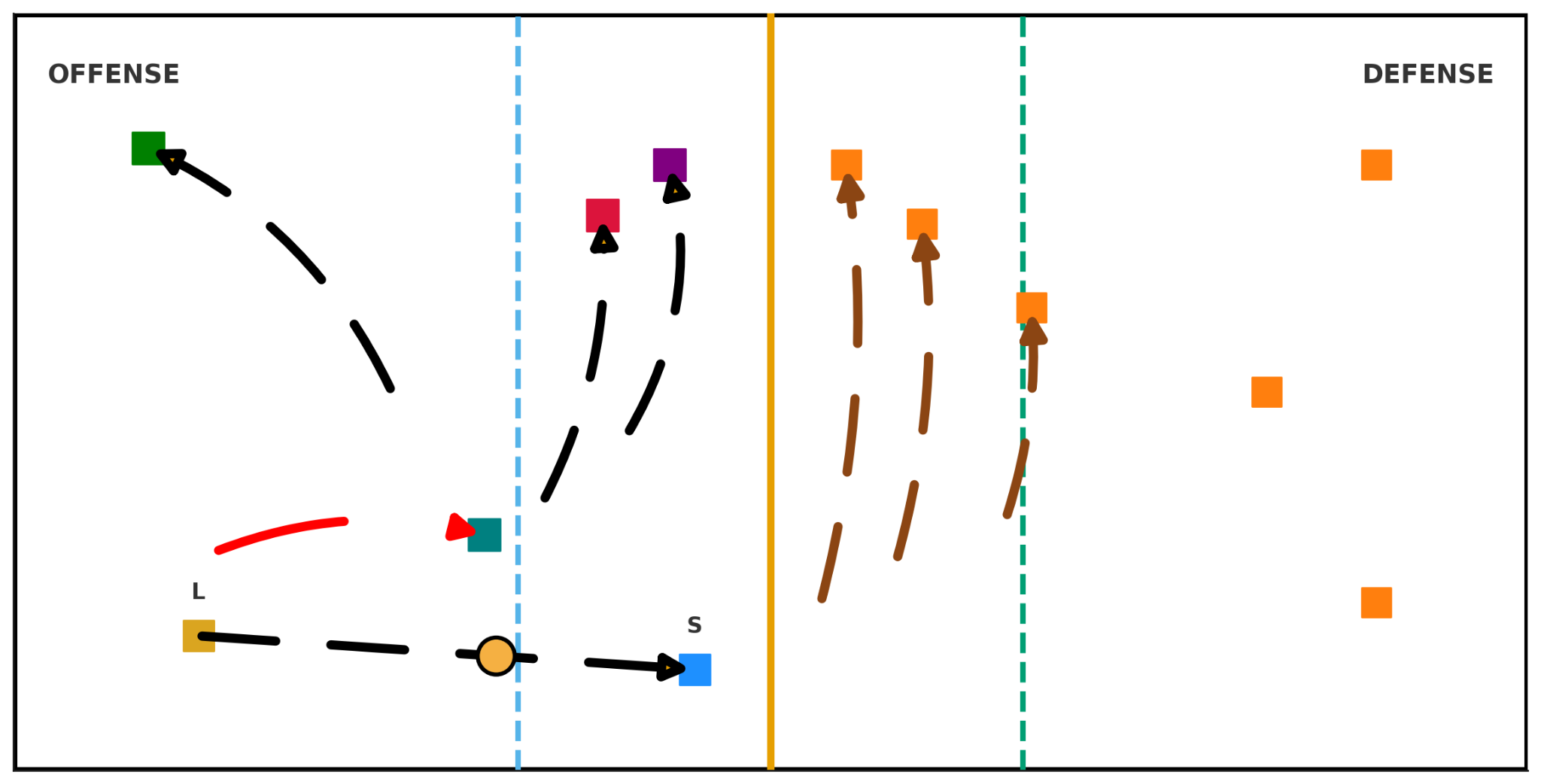}
  \caption{Volleyball coach schematic of the role-free \(3\text{–}1\) ``left–right'' split strategy, disguising the location of the set and spike attack. The runners aim to stretch the defensive formation toward a false corridor and create a viable lane for the true attack.}
  \label{left_right.png}
\end{figure}

Next, we consider a multi-agent deception scenario in volleyball, modeling the court as a rectangular grid $(\mathcal{V}, \mathcal{E})$. To disguise the setup and location of a spike shot close to the net, the team coordinates its movement before the shot using a ``$3\text{--}1$ split'' strategy across the left and right 
attack corridors. 
{Figure~\ref{left_right.png} provides a coach's schematic of this role-free $3\text{--}1$ split strategy, in which three runners threaten the false corridor while one maintains a viable lane toward the true attack corridor.}
Three attackers move to threaten one corridor to draw the defensive blockers, while a fourth attacker maintains a viable lane in the opposite corridor. This tactical misdirection is designed to force the defense to account for multiple approach vectors to the net simultaneously. 

The offense consists of 
a designated setter $a_s$ and runners $\mathcal{R} = \{1, \dots, N\} \setminus \{a_s\}$. During the offensive sequence, the team aims to converge upon a true attack region $\mathcal{K}_T \subset \mathcal{V}$ while feigning an approach toward a false region $\mathcal{K}_F \neq \mathcal{K}_T$. To realize this team-level intent, we specialize the per-step cost $\mathcal{C}_t$ to include both individual target-tracking and tactical coupling terms, expressed as
\begin{equation}
\label{eq:Ct_volleyball_full}
\mathcal{C}_t = \sum_{i \in \mathcal{I}} d(s_i(t+1), s_i^{\text{ref}}) + \boldsymbol{\phi}_{\mathrm{split}}(t) + \boldsymbol{\phi}_{\mathrm{space}}(t)
\end{equation}
where $s_i^{\text{ref}}$ is a transient reference state assigned dynamically at each step based on the runner's proximity to the assigned tactical roles. 

\vspace{-12pt}

We divide the attacking side into two corridors $\sigma: \mathcal{V} \to \{-1, +1\}$, and let \linebreak  $\sigma_F,\sigma_T\in\{-1,+1\}$ denote the labels associated with the false and true attack corridors, respectively. The number of runners in each corridor is given by 
\begin{equation}
\label{eq:sidecounts}
\begin{aligned}
S_F(t) &= \sum_{i\in\mathcal{R}} \tfrac{1}{2}
\left[1+\sigma(s_i(t))\sigma_F\right], \\
S_T(t) &= \sum_{i\in\mathcal{R}} \tfrac{1}{2}
\left[1+\sigma(s_i(t))\sigma_T\right].
\end{aligned}
\end{equation}
The role-free coupling $\boldsymbol{\phi}_{\mathrm{split}}(t)$ then softly induces the $3\text{--}1$ distribution by penalizing deviations from this target ratio,
\begin{equation}
\label{eq:phi-split}
\boldsymbol{\phi}_{\mathrm{split}}(t) = \gamma_F(S_F(t)-3)^2 + \gamma_T(S_T(t)-1)^2
\end{equation}
where we set the coupling gains to $\gamma_F = \gamma_T = 12.0$. To avoid bunching and path crossings, we add a short-range repulsion term,
\begin{equation}
\label{eq:phi-space}
\boldsymbol{\phi}_{\mathrm{space}}(t) = \sum_{\{i,j\}\subseteq\mathcal{R}} \frac{\gamma_{\mathrm{sep}}}{\epsilon + d(s_i(t), s_j(t))}
\end{equation}
where $\gamma_{\mathrm{sep}} = 0.5$ and $\epsilon = 0.1$ are fixed constants. 
Substituting the task-specific cost $\mathcal{C}_t$ from \eqref{eq:Ct_volleyball_full} into the joint-cost framework of \eqref{eq:boltzmann_distribution_compute} with $\lambda=6.5$ yields the PMF for our stochastic policy. By sampling joint actions from the resulting PMF accordingly, the runners' trajectories emerge stochastically to satisfy the $3\text{--}1$ split while adapting to real-time positions.

{\textls[-15]{We run this implementation for 500 trials. Each trial rollout consists of $T=22$ steps,} with the resulting trajectories and visit-frequency heatmap shown in Figure~\ref{fig:left-right-heatmaps}. The split cost in \eqref{eq:phi-space} favors joint configurations with three runners in the false corridor and one runner in the true corridor, without prescribing which runner occupies each role. Consequently, the runner maintaining the true-side attack lane can vary according to the current team configuration. Figure~\ref{fig:left-right-heatmaps} illustrates this behavior for both choices of the false corridor; when the false corridor is switched between the upper and lower sides, the concentration of runner trajectories shifts accordingly while a viable attack lane is maintained on the opposite side. These results demonstrate that the role-free coupling induces the desired\linebreak   $3\text{--}1$ deceptive split under the joint stochastic policy.
}

\begin{figure}[H] 
  
  \subfloat[Upper false corridor tactic ($G_F = \text{UP}$).\label{fig:heatmap-up}]{
    \includegraphics[width=0.8\columnwidth]{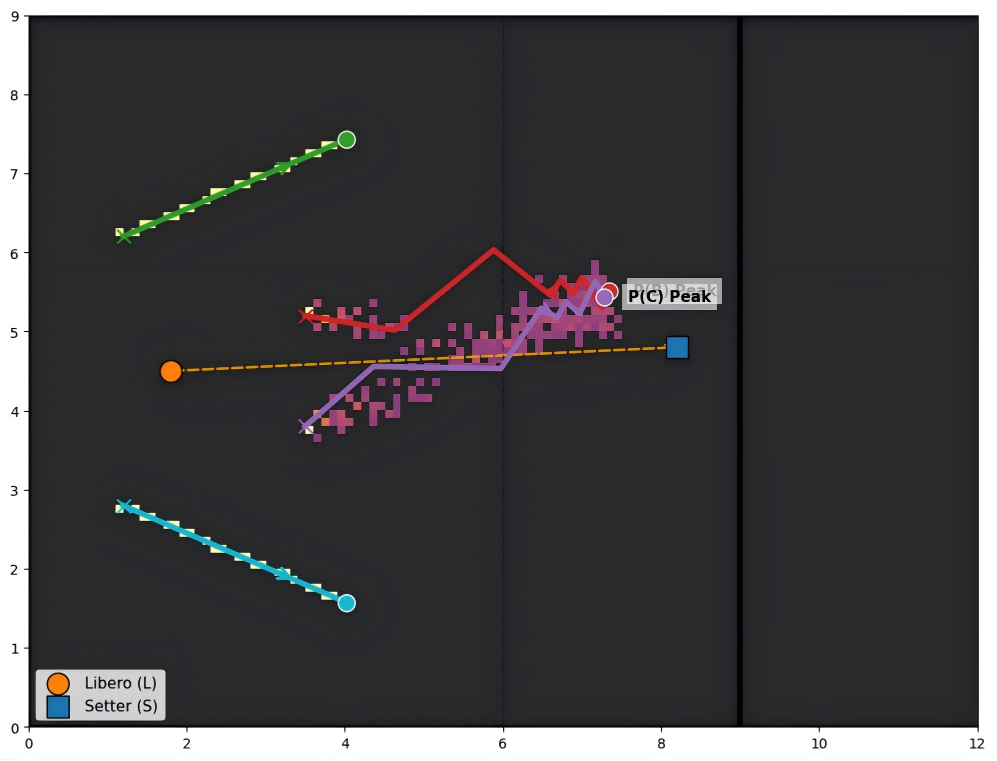}
  } \\ 
  
  \vspace{0.2em} 

\end{figure}

\begin{figure}[H]\ContinuedFloat
%

  \subfloat[Lower false corridor tactic ($G_F = \text{DOWN}$).\label{fig:heatmap-down}]{
    \includegraphics[width=0.8\columnwidth]{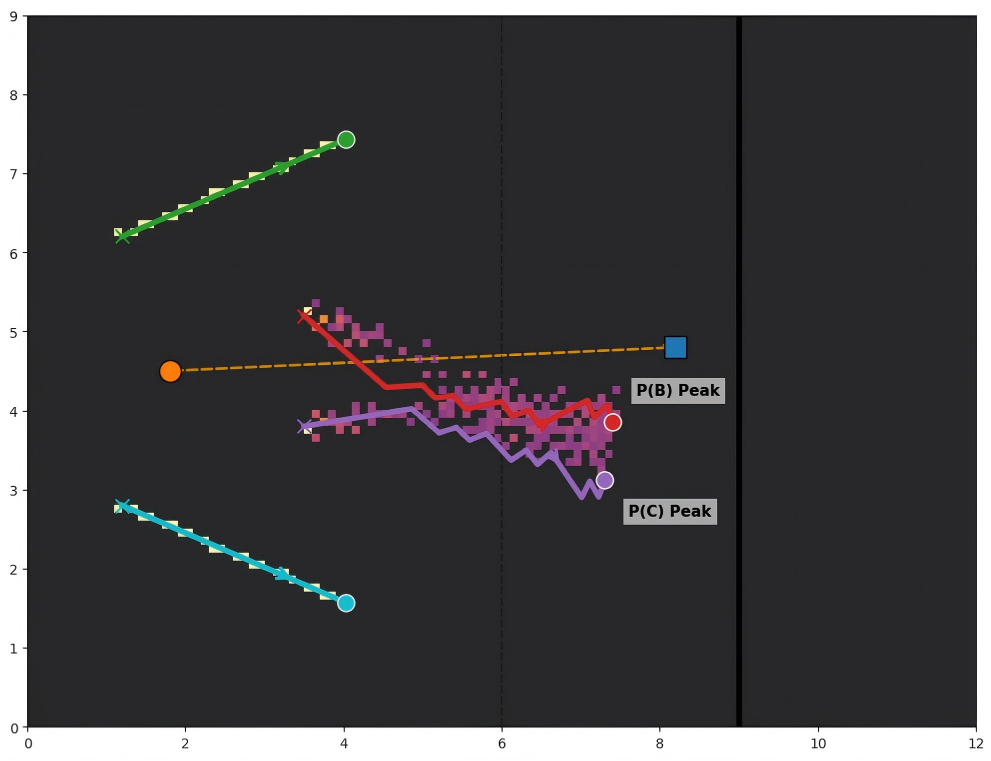}
  }

  \caption{\textls[-25]{Volleyball example of the deceptive split over 300 rollouts. The heatmaps demonstrate the emergent 3--1 coordinated misdirection: (\textbf{a}) three runners bias toward the upper false side; (\textbf{b}) three runners} bias toward the lower side, while one preserves a viable true-side attack lane.}
  \label{fig:left-right-heatmaps}
\end{figure}

\section{Conclusions and Future Work}\label{CL}

This paper introduced a unified deceptive path planning framework that uses a Boltzmann distribution over short-horizon trajectory candidates within a receding-horizon loop. 
This approach relies on local position and constraint sensing with deterministic cost and produces stochastic trajectories that preserve deception without training.
Our method introduces and enables new forms of dynamic deception planning and extends naturally from single- to multi-agent settings via parameter sharing through coupled cost terms. 
Step-by-step tuning enables new forms of dynamic deception and time-varying cost, adapting to the local environment and incorporating constraints such as an energy budget. 
This permits, for example, moving goals and false goals, goal reassignment, role swapping among agents, changing the form of deception, and tuning the desired level of deception. 
Similar to other deceptive path planners, the work so far has only modeled a passive adversarial observer. Nevertheless, our framework opens the door to incorporating adversarial agents. For example, the use of moving true and false goals can be applied to deceptive pursuit-evasion. 

It is of interest to further develop decentralized and partially observed variants that limit communications while preserving coordinated team deception (for example, using the common model of restricting to nearest neighbor communications). 
It is also of interest to introduce perception, networking, and new forms of dynamic deception. For example, adaptive deception may be linked to perception of adversarial agents. 
Learning cost weights and deception features from data (e.g., human demonstrations or adversarial self-play) may broaden applicability. 
Finally, we believe the method can be more generally applied in other dynamic deceptive resource allocation settings.

\section*{Funding}
This work was supported in part by the Office of Naval Research (ONR)
under grant number N00014-22-1-2626.

\end{document}